\documentclass[Journal]{IEEEtran}

\IEEEoverridecommandlockouts

\usepackage{graphics} 
\usepackage{graphicx}
\usepackage{epsfig} 
\usepackage{mathptmx} 
\usepackage{times} 
\usepackage{amsmath} 
\usepackage{amssymb}  
\usepackage{amsfonts}
\usepackage{float}
\usepackage{cite}
\usepackage{subfigure}
\usepackage[latin9]{inputenc}
\usepackage{booktabs}
\usepackage{graphicx}
\usepackage{algorithm}
\usepackage{algorithmicx}
\usepackage{algpseudocode}
\usepackage{setspace}
\usepackage{comment}
\usepackage{color}
\usepackage{url}
\usepackage{multirow}
\usepackage{threeparttable}
\usepackage{multirow}

\makeatletter

\newcommand{\Rmnum}[1]{\expandafter\@slowromancap\romannumeral #1@}

\newenvironment{breakablealgorithm}
  {
   \begin{center}
     \refstepcounter{algorithm}
     \hrule height.8pt depth0pt \kern2pt
     \renewcommand{\caption}[2][\relax]{
       {\raggedright\textbf{\ALG@name~\thealgorithm} ##2\par}%
       \ifx\relax##1\relax 
         \addcontentsline{loa}{algorithm}{\protect\numberline{\thealgorithm}##2}%
       \else 
         \addcontentsline{loa}{algorithm}{\protect\numberline{\thealgorithm}##1}%
       \fi
       \kern2pt\hrule\kern2pt
     }
  }{
     \kern2pt\hrule\relax
   \end{center}
  }
\makeatother

\begin{document}

\title{Online Scheduling of a Residential Microgrid via Monte-Carlo Tree Search and a Learned Model}

\author{Hang Shuai,~\IEEEmembership{Member, ~IEEE}, Haibo He,~\IEEEmembership{Fellow,~IEEE}
\vspace{-0.8cm}

\thanks{H. Shuai, and H. He are with the Department of Electrical, Computer and Biomedical Engineering, University of Rhode Island, Kingston, RI, USA.}


\thanks{\copyright 20XX IEEE.  Personal use of this material is permitted.  Permission from IEEE must be obtained for all other uses, in any current or future media, including reprinting/republishing this material for advertising or promotional purposes, creating new collective works, for resale or redistribution to servers or lists, or reuse of any copyrighted component of this work in other works.}



}
\markboth{SUBMITTED TO IEEE FOR POSSIBLE PUBLICATION. COPYRIGHT WILL BE TRANSFERRED WITHOUT NOTICE.}%
{Shell \MakeLowercase{\textit{et al.}}: Bare Demo of IEEEtran.cls for IEEE Transactions on Magnetics Journals}

\date{}

\maketitle
\begin{abstract}
The uncertainty of distributed renewable energy brings significant challenges to economic operation of microgrids.
Conventional online optimization approaches require a forecast model.
However, accurately forecasting the renewable power generations is still a tough task.
To achieve online scheduling of a residential microgrid (RM) that does not need a forecast model to predict the future PV/wind and load power sequences, this paper investigates the usage of reinforcement learning (RL) approach to tackle this challenge. Specifically, based on the recent development of $\textbf{M}$odel-Based $\textbf{R}$einforcement $\textbf{L}$earning, \textit{MuZero}, we investigate its application to the RM scheduling problem. 
To accommodate the characteristics of the RM scheduling application, a optimization framework that combines the model-based RL agent with the mathematical optimization technique is designed, and long short-term memory (LSTM) units are adopted to extract features from the past renewable generation and load sequences.
At each time step, the optimal decision is obtained by conducting Monte-Carlo tree search (MCTS) with a learned model and solving an optimal power flow sub-problem.
In this way, this approach can sequentially make operational decisions online without relying on a forecast model.
The numerical simulation results demonstrate the effectiveness of the proposed algorithm.

\end{abstract}


\section*{Nomenclature}
\addcontentsline{toc}{section}{Nomenclature}
\begin{IEEEdescription}[\IEEEusemathlabelsep\IEEEsetlabelwidth{$V_1,V_2,V_3,V_4$}]
\setlength{\parskip}{2pt}
\item[$\textbf{Superscript, Subscript, Indices and Sets}$]
\item[$b, grid$] Battery and utility grid.
\item[$ch, dis$] Charge and discharge mode of battery.
\item[$g, \mathcal{G}$] Index and set of dispatchable generators.
\item[$K, k$] Number and index of hypothetical time steps.
\item[$L$] Load power demand.
\item[$T, t, \Gamma$] Number, index, and set of time steps.
\item[$i, j, \mathcal{N}$] Index and set of buses.
\item[$\Upsilon$] Set of branches in microgrid.
\item[$max, min$] Maximum and minimum value.
\item[$wt, pv$] Wind turbine and PV panel.

\item[$\textbf{Variables}$]
\item[$P, Q$] Active power and reactive power.
\item[$SoC$] State of charge of battery.
\item[$l_{ij}, v_{i}$] Square of branch current and bus voltage.
\item[$I^{ch}, I^{dis}$] Charge and discharge state of battery.
\item[$s, x, r$] State, decision, and reward of microgrid.
\item[$\hat s, \hat x, \hat r$] Internal state, decision, and reward of microgrid.

\item[$\textbf{Parameters}$]
\item[$p_{cur}$] Renewable energy curtailment cost coefficient.
\item[$r_{ij}, x_{ij}$] Resistance and reactance of power line.
\item[$\alpha_g, \beta_g, c_g$] Fuel cost coefficient of dispatchable DGs.
\item[$\rho$] Unit degradation price of battery.
\item[$\eta$] Efficiency of battery.
\item[$E^{min}, E^{max}$] Minimum and maximum stored energy in battery.
\item[$\gamma$] Discout factor.
\item[$\Delta t$] Time resolution.

\item[$\textbf{Functions}$]
\item[$C(\cdot), r(\cdot)$] Cost function and reward function.
\item[$h_{\theta}(\cdot), g_{\vartheta}(\cdot)$] Representation and dynamic network.
\item[$ f_{\phi}(\cdot)$] Prediction network.
\item[$ l(\cdot)$] Loss network.
\item[$ Q(\hat s, \hat x)$] Mean value function.
\item[$ P(\hat s, \hat x)$] Policy function.
\item[$ S(\hat s, \hat x)$] Transition function.

\end{IEEEdescription}

\section{Introduction} \label{Introduction}
Microgrid is a group of interconnected loads and distributed energy resources within clearly defined electrical boundaries that acts as a single controllable entity with respect to the grid \cite{ton2012us}.
It is becoming a widely adopted technology to utilize distributed energy resources (DERs) as their capability of reducing greenhouse emissions, improving consumers' supply reliability, enhancing power grid resiliency, etc., \cite{7120901,giraldez2018phase}.
For instance, there have been 6610 microgrid projects representing 31.7 GW of planned and installed power capacity globally \cite{Navigant}, as of March 2020.
However, the intermittent and uncertainty of integrated renewable energy bring significant challenges to the reliable and economic operation of microgrids.
The optimal optimization and control strategies are the key techniques to ensure the economic operation of microgrids.
As a result, the optimization of microgrids has obtained extensive research, and a variety of microgrid energy management algorithms (see \cite{zia2018microgrids} and the references therein) have been proposed to deal with the uncertainties, such as, linear and nonlinear programming methods, dynamic programming and rule-based methods, meta-heuristic approaches (particle swarm optimization, genetic algorithm, etc.), artificial intelligence methods (fuzzy logic, neural network, multi-agent system, etc.), stochastic programming, and robust optimization approaches \cite{Lei2019Robust}.
 
However, the above microgrid optimization methods are mainly proposed to solve the microgrid planning problems (see \cite{Alharbi2018Stochastic,Lei2019Robust,Yuan2017Co}) or day-ahead scheduling problems (see \cite{liuBidding2016,Farzin2017A}).
Using these methods, we can make day-ahead scheduling according to the forecast information of all the future system state and the statistic distribution information of the uncertainties in the system.
Since the prediction errors come from both the power generation side and the demand side, the actual operational decisions are re-optimized sequentially in the intra-day online optimization process according to the updated short-term forecasting information provided by the forecast model of the system. 
Model predictive control (MPC) is a traditional online optimization method that has been applied in microgrids \cite{wang2014control,Yang2019RealTime}.
But, the performance of the MPC approach depends on the precision of the forecasting generation/load power and the fitted statistic distribution information.
To reduce the influence of renewable energy forecast errors on the economic operation of microgrids, Hang \textit{et al}. \cite{Shuai2019On} proposed a cost function approximation (CFA) based online optimization algorithm.
However, the CFA based online optimization strategy is still rely on the short-term renewable energy forecast information.

To obtain optimal online operation decisions, researchers have made some efforts and proposed several optimization approaches to reduce the dependence on forecasting information provided by forecast models of microgrids.
A class of heuristic online algorithms, called CHASE \cite{lu2013online}, \textit{h}CHASE \cite{jia2019retroactive,jia2020novel}, are proposed recently.
The CHASE algorithm can achieve theoretical performance guarantee without any future information.
However, the CHASE algorithm is designed to solve single or multiple homogeneous local generators.
Then, a more general retrospection-inspired online scheduling algorithm, \textit{h}CHASE, is proposed.
Shi \textit{et al}. \cite{shi2015real} proposed a Lyapunov optimization based online optimization strategy for microgrids, and simulations demonstrate that the algorithm can make sub-optimal decisions without any a prior statistical knowledge of the stochastic processes.
Wann-Jiun \textit{et al}. \cite{ma2016distributed} proposed an online alternating direction method of multipliers (ADMM) based distributed algorithm for online optimization of microgrids.
The online ADMM algorithm proposed in \cite{ma2016distributed} does not require any forecast data to proceed, which avoids problems caused by inaccurate forecasting.
However, simulation results in \cite{ma2016distributed} indicate that there exists constraint violations. More concretely, the online ADMM algorithm cannot ensure the active/reactive power constraints and the voltage constraints be fulfilled at each time period.
Li \textit{et al}. \cite{LiRealTime2019} proposed an online learning-aided energy management algorithm and combined it with ADMM algorithm to facilitate the real-time implementation.
Rahbar \textit{et al}. \cite{rahbar2014real} developed dynamic programming (DP) based sequential online optimization algorithm.
The authors of this paper proposed an approximate dynamic programming (ADP) based microgrid online optimization approaches \cite{Hang2019TSG,Hang2019TSTE}.
After trained off-line using the day-ahead forecasting information, the ADP algorithm can obtain near-optimal online decisions only according to the current system state and the well-trained value functions.

Although the above online optimization approaches reduced the dependence on intra-day forecasting information, the historical renewable and load power data are not been fully utilized in the optimization process.
With the rapid development of machine learning techniques, researchers have made some efforts to apply deep learning and model-free reinfrocement learning algorithms to solve microgrid energy management problem \cite{Du8769895}.
Besides, prior research works \cite{gao2014machine,Lei2020DynamicED} indicate that the intelligent agent trained based on historical data is more general to adapt to the unknown situation in the future.
Thus, in order to learn to operate microgrids from historical data, researchers proposed the model-free deep reinforcement learning (DRL) based online optimization algorithms \cite{franccois2016deep,ji2019real,Lei2020DynamicED} recently.
In \cite{ji2019real,Lei2020DynamicED}, the historical data are used to train the designed Deep Q Network (DQN) and Deep Deterministic Policy Gradient (DDPG) algorithms to achieve a good online decision performance.

In this work, we investigate the application of a model-based deep reinforcement learning (MB-DRL) algorithm developed by reference \cite{schrittwieser2019mastering} to solve the microgrid scheduling problem.
Different from the \textit{AlphaGo} \cite{silver2016mastering} and \textit{AlphaGo Zero} \cite{silver2017mastering} algorithms that need to know the dynamics of the environment, the MB-DRL algorithm, called \textit{MuZero}, that proposed in \cite{schrittwieser2019mastering} is a more general and more powerful reinforcement learning algorithm that can achieve superhuman performance in a range of challenging Atari games. 
Specifically, the \textit{MuZero} algorithm combines a tree-based search policy with a learned model that consists of three networks (representation network, dynamic network, and prediction network), and can make decisions without any knowledge of the underlying dynamics of the environment.
It is worth noting that there are many differences between the microgrid optimization problem and playing Atari games.
For example, there are plenty of equality and inequality constraints in microgrids, and the action space of the optimization problem in this work is huge because the decision variables are multidimensional.
These differences bring significant challenges to the application of the algorithm.
This motivated us to explore the application potential of the MB-DRL algorithm in solving microgrid online optimization problems.

This work focuses on the online optimization of a residential microgrid (RM) under uncertainty.
The optimization problem in this paper is formulated as a mixed integer second-order cone programming (MISOCP) problem.
To solve this problem, we reformulate it as a Markov Decision Process (MDP), and a MB-DRL based RM scheduling algorithm is designed.
The advantage of the developed algorithm is that it can make online decisions sequentially without relying on the renewable and load power prediction from forecast models.
The main contributions of this work are summarized as follows:
\begin{itemize}
\item[1)] An MB-DRL based RM optimization approach is developed based on \textit{MuZero} \cite{schrittwieser2019mastering} algorithm, with the aim of conceiving an online optimization agent that can teach itself how to optimally operate the RM system through self-play.
\item[2)] To deal with the application challenges brought by a large number of operational constraints and the huge decision space, a framework that combines the MB-DRL algorithm with mathematical optimization techniques is designed.
\item[3)] A new representation network architecture is designed.
Different from most Atari games existing large spatial resolution in observations, the observations of microgrids have a strong time correlation.
Thus, for the representation network in the proposed scheduling algorithm, we first utilize three LSTM units to extract features from past solar power, wind power, and load power, respectively. Then, the feature vectors are concatenated with the current state information and fed into the input layer of a fully connected neural network.
\item[4)] Simulation results demonstrate the developed MB-DRL approach can learn to solve the complex RM optimization problem by self-play, and can obtain better online optimization performance than many state-of-the-art online optimization algorithms.
\end{itemize}

The rest of this paper is organized as follows. The online scheduling of the residential microgrid is formulated in Section II. Section III presents the MB-DRL based online optimization algorithm. The simulation results are given in Section IV. Section V concludes the paper.

\vspace{-0.5em}
\section{Scheduling Model of the Residential Microgrid} \label{Real-Time Scheduling Model of the Residential Microgrid}
The microgrid system investigated in this paper is a RM system with a high penetration of renewable energy, which consists of PV-based DER units, wind turbine based DER units, diesel engine generator, energy storage device, electrical loads, and a smart energy management unit.
The microgrid is connected with the utility grid, so it can purchase power from the utility grid when the RM suffers from power supply shortage.
On the other hand, the microgrid can also sell its surplus energy to the utility grid.
As the uncertainties from both renewable energy resources and the demand side, an energy management system (EMS) is needed to coordinate all the generation and load resources (including the energy storage system) to ensure the secure and economic operation of the system.
So, one of the functions of smart EMS unit is to make online energy scheduling decisions according to the actual load demand and the available generation capacity from all power sources.
In this paper, we investigate the designation of a smart online optimization algorithm to achieve the optimal operation of the system.  
In the following section, we formulate the optimization model of the RM system as a MISOCP problem.

From the perspective of a system operator, the objective of the online energy management is to make decisions at each time step in order to minimize the operational cost under the uncertainties from renewable energy and electricity load, which leads to the following economic dispatch model:
\begin{equation}\label{EQ1}
\begin{aligned}
\min \Xi \bigg\{ \sum_{t=\Delta t}^{T} \bigg( \sum_{g \in \mathcal{G}} C_g^{DG}\big(P_g^{DG}(t)\big) + C_{grid}\big(P_{buy}^{grid}(t), P_{sell}^{grid}(t)\big) \\ 
   + C_{bat}\big(P^{b}(t)\big) + C_{cur}\big( P_{cur}^{ren}(t)\big) \bigg) \bigg\}
\end{aligned}
\end{equation}
\begin{equation}\label{EQ2}
C_g^{DG}\big(P_g^{DG}(t)\big) = \big(\alpha_g (P_g^{DG}(t))^2 + \beta_g P_g^{DG}(t) + c_g\big) \Delta t  
\end{equation}
\begin{equation}\label{EQ3}
C_{grid}\big(P_{buy}^{grid}(t), P_{sell}^{grid}(t)\big) =  \big(p_{buy}(t)P_{buy}^{grid}(t) - p_{sell}(t)P_{sell}^{grid}(t)\big) \Delta t
\end{equation}
\begin{equation}\label{EQ4}
C_{bat}\big(P^{b}(t)\big) =  \rho \vert SoC(t) - SoC(t-\Delta t) \vert
\end{equation}
\begin{equation}\label{EQ5}
C_{cur}\big(P_{cur}^{ren}(t) \big) =  p_{cur} \big( \bar{P}^{pv}(t) + \bar{P}^{wt}(t) - P^{pv}(t) - P^{wt}(t) \big) \Delta t
\end{equation}
where $\Xi\{\cdot\}$ represents the expectational operator.
The first term in (\ref{EQ1}) represents the fuel cost of all dispatchable generators during a single time period, which is a quadratic function of the active power generation $P_g^{DG}(t)$, as shown in (\ref{EQ2}).
The second term of the objective function is the power exchange cost which can be calculated as shown in (\ref{EQ3}), where $P_{buy}^{grid}(t)$ and $P_{sell}^{grid}(t)$ are respectively power purchased and sold by the RM, and $p_{buy}(t)$ and $p_{sell}(t)$ are energy purchase and sell price, respectively.
Note that $p_{sell}(t)$ is usually lower than or equal to $p_{buy}(t)$, which is determined by electricity price policies in different states.
The third term in (\ref{EQ1}) denotes the degradation cost of the battery system.
The degradation cost caused by charging and discharging can be linearly approximated by the change between two consecutive SoC \cite{Du8769895,LiuDistributed2018}, as shown in (\ref{EQ4}).
The last term is the renewable energy curtailment cost of the system, where $P^{pv}(t)$ and $P^{wt}(t)$ are respectively the dispatched power of PV energy and wind energy.
$\bar{P}^{pv}(t)$ and $\bar{P}^{wt}(t)$ are respectively the maximum available PV power and wind power currently.
In the above equations, $t$ is the time period index.

\vspace{-0.2em}
The RM system also need to satisfy the following operational constraints:
\begin{equation}\label{EQ6}
P_g^{DG,min} \le P_g^{DG}(t) \le P_g^{DG,max}, \forall t \in \Gamma, \forall g \in \mathcal{G}
\end{equation}
\begin{equation}\label{EQ7}
(P_g^{DG}(t))^2 + (Q_g^{DG}(t))^2 \le (S_g^{DG,max})^2, \forall t \in \Gamma, \forall g \in \mathcal{G}
\end{equation}
\begin{equation}\label{EQ8}
0 \le P^{wt}(t) \le \bar{P}^{wt}(t), \forall t \in \Gamma
\end{equation}
\begin{equation}\label{EQ9}
(P^{wt}(t))^2 + (Q^{wt}(t))^2 \le (S^{wt,max})^2, \forall t \in \Gamma
\end{equation}
\begin{equation}\label{EQ10}
0 \le P^{pv}(t) \le \bar{P}^{pv}(t), \forall t \in \Gamma
\end{equation}
\begin{equation}\label{EQ11}
(P^{pv}(t))^2 + (Q^{pv}(t))^2 \le (S^{pv,max})^2, \forall t \in \Gamma
\end{equation}
\begin{equation}\label{EQ12}
\left\{ 
   \begin{array}{lr} 
   0 \leq P_{buy}^{grid}(t) \leq P_{buy}^{grid,max} \\
   0 \leq P_{sell}^{grid}(t) \leq P_{sell}^{grid,max}
   \end{array}
    \forall t \in \Gamma
\right.  
\end{equation}
\begin{equation}\label{EQ13}
   0 \leq Q^{grid}(t) \leq Q^{grid,max}, \forall t \in \Gamma 
\end{equation}
\begin{equation}\label{EQ14}
\left\{ 
   \begin{array}{lr} 
   0 \leq P^{ch}(t) \leq I^{ch} (t)P^{ch,max} \\
   0 \leq P^{dis}(t) \leq I^{dis} (t) P^{dis,max}
   \end{array}
    \forall t \in \Gamma
\right.  
\end{equation}
\begin{equation}\label{EQ15}
\begin {aligned}
   P^{b} (t) = I^{dis} (t) P^{dis} (t) - I^{ch} (t) P^{ch} (t), \forall t \in \Gamma
\end {aligned}
\end{equation}
\begin{equation}\label{EQ16}
\begin {aligned}
   I^{dis} (t) + I^{ch} (t) \le 1, \forall t \in \Gamma, \{I^{dis} (t), I^{ch} (t)\} \in \{0 ,1\}
\end {aligned}
\end{equation}
\begin{equation}\label{EQ17}
(P^{b}(t))^2 + (Q^{b}(t))^2 \le (S^{b,max})^2, \forall t \in \Gamma
\end{equation}
\begin{equation}\label{EQ18}
   SoC(t+\Delta t) = SoC (t) + \eta^{ch} \frac{P^{ch} (t)}{E^{max}} \Delta t - \frac{1}{\eta^{dis}} \frac{P^{dis}(t)}{E^{max}} \Delta t
\end{equation}
\begin{equation}\label{EQ19}
   SoC^{min} \leq SoC (t) \leq SoC^{max}, \forall t \in \Gamma
\end{equation}
\begin{equation}\label{EQ20}
\left\{ 
   \begin{array}{lr} 
   P_j(t) = P_{ij}(t) - r_{ij} \textit{l}_{ij}(t) - \sum_{m:(j,m)\in \Upsilon} P_{jm} (t)\\
   Q_j(t) = Q_{ij}(t) - x_{ij} \textit{l}_{ij}(t) - \sum_{m:(j,m)\in \Upsilon} Q_{jm} (t)\\
   \end{array}
    \forall (i, j) \in \Upsilon, \forall t \in \Gamma
\right.  
\end{equation}
\begin{equation}\label{EQ21}
   v_j(t) = v_i(t) - 2 \big(r_{ij} P_{ij}(t) + \textit{x}_{ij}Q_{ij}(t)\big) + \big(r_{ij}^2 + x_{ij}^2\big) \textit{l}_{ij}(t), \forall t \in \Gamma
\end{equation}
\begin{equation}\label{EQ22}
   V_i^{min} \le \mid V_i(t) \mid \le V_i^{max}, \forall t \in \Gamma, \forall i \in \mathcal{N}
\end{equation}
\begin{equation}\label{EQ23}
   \textit{l}_{ij}(t) = \frac{P_{ij}(t)^2 + Q_{ij}(t)^2}{v_i(t)}, \forall (i, j) \in \Upsilon, \forall t \in \Gamma
\end{equation}
Constraints (\ref{EQ6})-(\ref{EQ7}) are the generator capacity constraints of controllable DGs, and (\ref{EQ7}) ensures that the reactive power generated at the inverter of controllable DGs will not exceed its capacity;
constraints (\ref{EQ8})-(\ref{EQ11}) are the generator capacity constraint of renewable generators, and the dispachable active generation power of the PV and wind turbine are limited by the current available renewable power $\bar{P}^{wt}(t)$ and $\bar{P}^{pv}(t)$, as shown in (\ref{EQ8}) and (\ref{EQ10});
constraints (\ref{EQ12})-(\ref{EQ13}) are the power exchange limitation between the RM and utility grid;
constraints (\ref{EQ14})-(\ref{EQ19}) are the battery storage related constraints, and constraint (\ref{EQ14}) ensures the active charge/discharge power will not exceed power limitations of the battery;
constraint (\ref{EQ16}) avoids simultaneous charging and discharging of batteries;
constraint (\ref{EQ17}) bounds the complex power of the battery, where $S^{b, max}$ is the capacity of the battery inverter;
constraint (\ref{EQ18}) is the energy transition constraint of the battery.

Constraints (\ref{EQ20})-(\ref{EQ23}) are the power flow model of the RM, where $P_{ij}(t)$ and $Q_{ij}(t)$ are the complex power flowing from bus $i$ to $j$.
In this work, the branch flow model \cite{LowConvex2014} is adopted to model the steady-state power flows in the power network, where $(i, j) \in \Upsilon$ denotes the branch between bus $i$ and $j$, and $r_{ij}+\textbf{i} x_{ij}$ is the complex impedance of the branch.
In (\ref{EQ20}), the complex net load of each bus $i$ is represented by $P_i(t)+\textbf{i}Q_i(t)$, which is the load power $P_i^L(t)+\textbf{i}Q_i^L(t)$ minus the generation power.
In the power flow model $\textit{l}_{ij}(t) = \mid I_{ij}(t)\mid^2$ and $v_{i}(t) = \mid V_{i}(t)\mid^2$, where $I_{ij}(t)$ and $V_{i}(t)$ are the branch current and complex voltage at each bus, respectively.
The magnitude of bus voltage is limited by (\ref{EQ22}).
To formulate the optimization problem as a convex problem, we relax the quadratic equality constraint (\ref{EQ23}) to the following inequality
\begin{equation}\label{EQ24}
   \textit{l}_{ij}(t) \ge \frac{P_{ij}(t)^2 + Q_{ij}(t)^2}{v_i(t)}, \forall (i, j) \in \Upsilon, \forall t \in \Gamma
\end{equation}

Finally, we can make the optimal decisions by solving the following MISOCP problem:  
\begin{equation}\label{EQ25}
\begin{aligned}
\min_{x_t} \quad & \Xi \bigg\{ \sum_{t=\Delta t}^{T} \bigg( \sum_{g \in \mathcal{G}} C_g^{DG}\big(P_g^{DG}(t)\big) + C_{grid}\big(P_{buy}^{grid}(t), P_{sell}^{grid}(t)\big) \\ 
   &+ C_{bat}\big(P^{b}(t)\big) + C_{cur}\big( P_{cur}^{ren}(t)\big) \bigg) \bigg\}
   \\
    \mbox{s.t.}\quad & (\ref{EQ6})-(\ref{EQ22}), \emph{and} \ (\ref{EQ24})
\end{aligned}
\end{equation}
where $x_t$ is the decision vector at time $t$, and 
\begin{equation}\label{EQ26}
\begin{aligned}
   x_t = \big(S_g^{DG}(t), S^{b}(t), S^{pv}(t), S^{wt}(t), P_{buy}^{grid}(t), P_{sell}^{grid}(t), Q^{grid}(t),\\
    P_{ij}(t), Q_{ij}(t), v_i(t), \textit{l}_{ij}(t)\big)
    \end{aligned}
\end{equation}
In (\ref{EQ26}), $S_g^{DG}(t), S^{pv}(t)$, and $S^{wt}(t)$ represent the complex power generation of controllable DGs, PV panels, and wind turbines, respectively.
$S^{b}(t)$ is the complex output power of the battery, and $S^{b}(t) = P^{b}(t) + \textbf{i} Q^{b}(t)$.

In this paper, we investigate the online optimization of the RM system, which means system operators need to make operational decisions according to the current state information of the RM.
In the following sections, we will introduce a data-driven method, namely the MB-DRL algorithm, to solve the online optimization problem.


\vspace{-0.5em}
\section{Microgrid Online Scheduling Via MCTS and a Learned Model}
In this section, a model-based deep reinforcement learning based RM online optimization approach is designed to optimally operate the RM under uncertainties from renewable energy and demand side.
We first reformulate the above RM optimization problem as a MDP problem.
Then, the adopted MB-DRL method is introduced.
Finally, a specific online optimization algorithm for the RM is designed.
\vspace{-0.5em}

\subsection{Problem Reformulation}
To facilitate the application of reinforcement learning approach, the optimization problem shown in (\ref{EQ26}) is reformulated as a MDP problem.
A MDP model includes some basic elements, namely state variables $s_t$, decision variables $x_t$, transition function, and reward function $r_t(\cdot)$, which will be defined in the following section.
Note that different from the variable representation rules in the previous equations, we will subscript $t$ to represent the variable at time $t$ in the following sections, which is to be consistent with the variable naming convention in the field of reinforcement learning.
For instance, the state variables is represented by $s_t$ rather than $s(t)$.
The state variables of the RM consist of the SoC of the battery, the active and reactive power demand of each bus $P_i^L(t)$ and $Q_i^L(t)$, the available PV generation $\bar P^{pv}(t)$, the available wind power generation $\bar P^{wt} (t)$, and the electricity price $p_{buy}^{grid} (t)$ and $p_{sell}^{grid}(t)$.
The state variables of the RM at time $t$ are defined below:
\begin{equation}\label{EQ27}
   s_t = \big\{SoC (t), P_i^L(t), Q_i^L(t), \bar P^{pv}(t), \bar P^{wt} (t), p_{buy}^{grid} (t), p_{sell}^{grid}(t) \big\}
\end{equation}

The decision variables of the RM at time $t$ is given in (\ref{EQ26}).
The SoC transition function has been given in (\ref{EQ18}).
The objective of the online optimization is to minimize the operational cost of the RM, which is equivalent to maximize the total rewards of the system.
So, we define the following reward function according to the objective function shown in (\ref{EQ1}):
\begin{equation}\label{EQ28}
\begin{aligned}
   r_t \big(s_t, x_t\big) = - \bigg( \sum_{g \in \mathcal{G}} C_g^{DG}\big(P_g^{DG}(t)\big) + C_{grid}\big(P_{buy}^{grid}(t), P_{sell}^{grid}(t)\big) \\ 
   + C_{bat}\big(P^{b}(t)\big) + C_{cur}\big( P_{cur}^{ren}(t)\big) \bigg)
\end{aligned}
\end{equation}
where, $r_t \big(s_t, x_t\big)$ is the reward of taking decision $x_t$ when RM in state $s_t$.

After reformulated the problem as a MDP problem, the next step is to design a reinforcement learning algorithm to make optimal decisions at each time step in order to maximize the cumulative rewards of the RM.

\vspace{-0.5em}
\subsection{The Model-Based Deep Reinforcement Learning Method} 
In this work, we adopted the MB-DRL optimization approach proposed in \cite{schrittwieser2019mastering}.
In this section, the principle of the the adopted MB-DRL approach is briefly introduced.
  
The MB-DRL algorithm proposed in \cite{schrittwieser2019mastering}, called \textit{MuZero}, combines Monte-Carlo Tree Search (MCTS) planning method with a learned neural network model.
Similar to traditional MCTS, the \textit{MuZero} algorithm involves iteratively building a search tree until some predefined computational budget (like a maximum iteration constraint) is reached, then the search is halted and the best action is determined according to the visit count of each action from the root node \cite{browne2012survey}.
\begin{figure}[t]\centering
\includegraphics[width=3.2in]{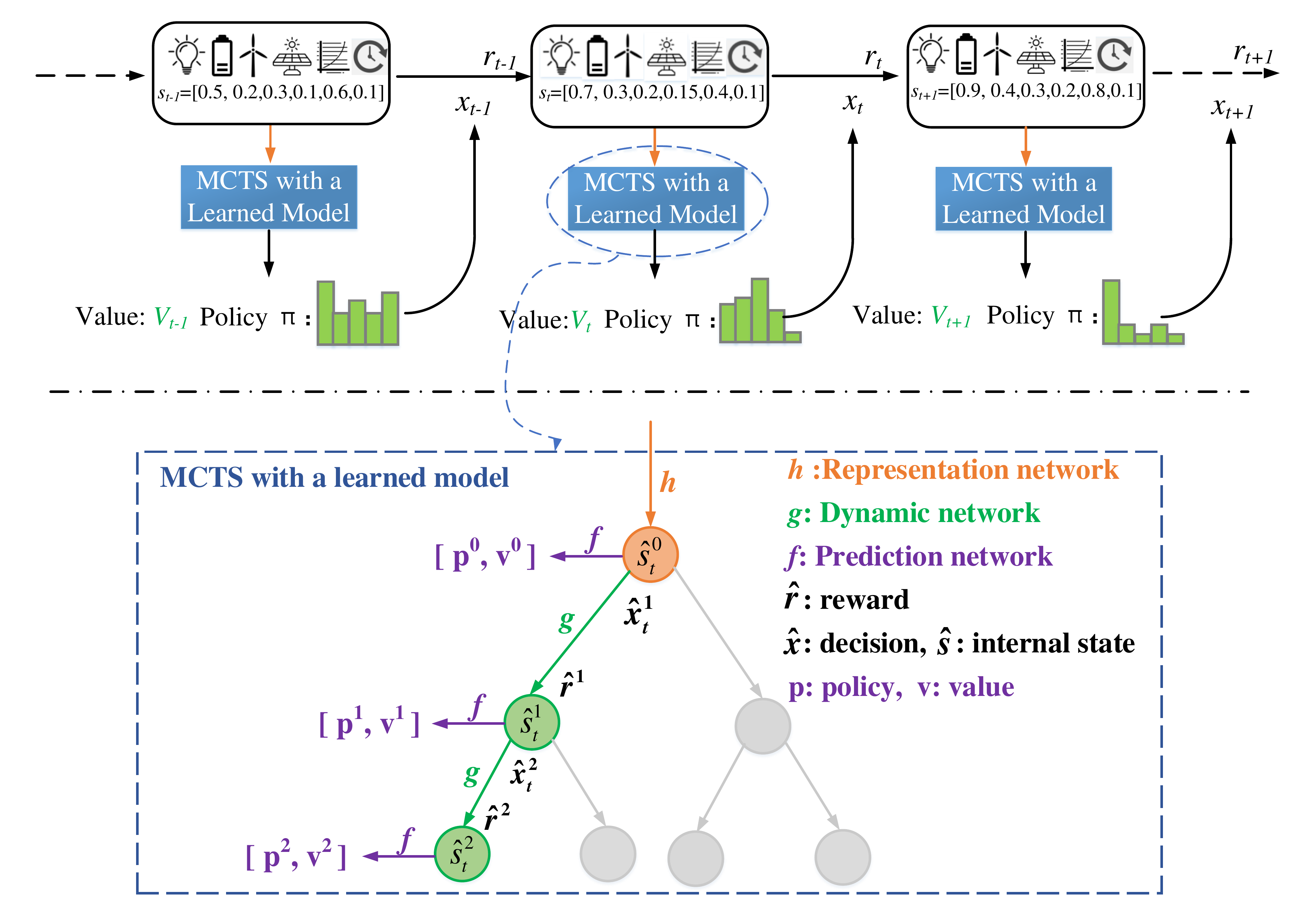}
\caption{The architecture of the MCTS with a learned model \cite{schrittwieser2019mastering}.} \label{fig:MicrogridZero-Architecture}
\vspace{-0.8em}
\end{figure}
The difference between the two algorithms is that a learned model is introduced in \textit{MuZero} to improve the performance of MCTS.
Specifically, three different neural networks, which constitute the learned model, are introduced in the tree search process, as shown in Fig. \ref{fig:MicrogridZero-Architecture}.
The first neural network is the representation network $h_{\theta}$, which is used to encode past system states (or observations) until the current time-step $t$, $(s_1,\cdots,s_t)$ , into an internal state $\hat s_t^0$.
The second one is the dynamic network $g_{\vartheta}$, which is used to get the next internal state $\hat s_t^{k}$ and the immediate reward $\hat r^k$ after taken an action $\hat x_t^k$ from an internal state $\hat s_t^{k-1}$.
The last neural network is the prediction network, which is used to get the control policy $\textbf{p}^k$ and value functions $v^k$ according to the internal state $\hat s_t^k$.
\begin{equation}\label{EQ29}
\begin {aligned}
   \hat s_t^0 = h_{\theta} (s_1,\cdots,s_t)
\end {aligned}
\end{equation}
\begin{equation}\label{EQ30}
\begin {aligned}
   \hat r^k, \hat s_t^k = g_{\vartheta}(\hat s_t^{k-1}, \hat x_t^k)
\end {aligned}
\end{equation}
\begin{equation}\label{EQ31}
\begin {aligned}
   \textbf{p}^k, v^k = f_{\phi}(\hat s_t^{k})
\end {aligned}
\end{equation}

In (\ref{EQ29})-(\ref{EQ31}), $\theta$ is the weights of the representation network;
$\vartheta$ is the weights of the dynamic network;
$\phi$ represents the weights of the prediction network.
The superscript $k$ represents the variables of the $k$th hypothetical time-step.
For instance, $\hat s_t^k$ denotes the internal state at the $k$th hypothetical time-step during the tree search conducted at the actual time-step $t$.
It is worth noting that the internal state has no semantics of the environment state attached to it \cite{schrittwieser2019mastering}.
The purpose of the internal state is to improve the prediction accuracy of the above control policies $\textbf{p}$, values $v$, and immediate rewards $\hat r$ that used in the tree search process.
We can also find that the dynamic network is actually an environment model approximator, which can directly compute the next internal state according to the current state and the executed action.
This makes the \textit{MuZero} algorithm belongs to a model based reinforcement learning algorithm.

The three neural network are trained off-line using the data generated by the self-play process of the \textit{MuZero} algorithm.
And after the neural network model has been well-trained, the optimal decisions of each time-step can be made forward through time according to the current state of the environment, as shown in Fig. \ref{fig:MicrogridZero-Architecture}.
At each time-step $t$, an MCTS is performed to build a search tree.
Note that an MCTS consists of a predefined number of simulations, and each simulation includes three steps, namely selection, expansion, and backpropagation.
The above learned model is utilized in each simulation to help us to build the search tree.
Then the best decision is sampled according to the visit count of the child nodes of the root node.
The decisions are applied to the environment, and the agent gets a new observation of system state $s_{t+1}$ and the actual reward $r_{t+1}$ from the environment.
The above procedure repeats until the end of the game.
Here we just provided a brief description of the MuZero algorithm, and refer the readers to reference \cite{schrittwieser2019mastering} for more details. 

\subsection{Proposed MB-DRL based Online Scheduling Algorithm for the Residential Microgrid}
Originally, the \textit{MuZero} algorithm is proposed to play games like Atari games and Go. 
In these games, the action is usually a single-dimensional variable and discrete values.
So the action space at a single time-step is relatively small.
For example, in Atari 2600 game Breakout, the action space is ['noop', 'fire', 'right', 'left', 'right fire', 'left fire'].
However, the actions of the optimization problem in this paper is a multidimensional continuous vector, as shown in (\ref{EQ26}).
Thus the action space is huge.
This makes us face the 'curse of dimensionality' when applying reinforcement learning method to solve our problem.
Besides, there are plenty of equality and inequality constraints in the above RM model, which is difficult to directly handle for the \textit{MuZero} algorithm. 
How to ensure the decisions given by the reinforcement learning agent fulfill the complex constraints (\ref{EQ6})-(\ref{EQ22}), and (\ref{EQ24}) is also a big challenge we face.

\begin{figure}[t]\centering
\includegraphics[width=3.2in]{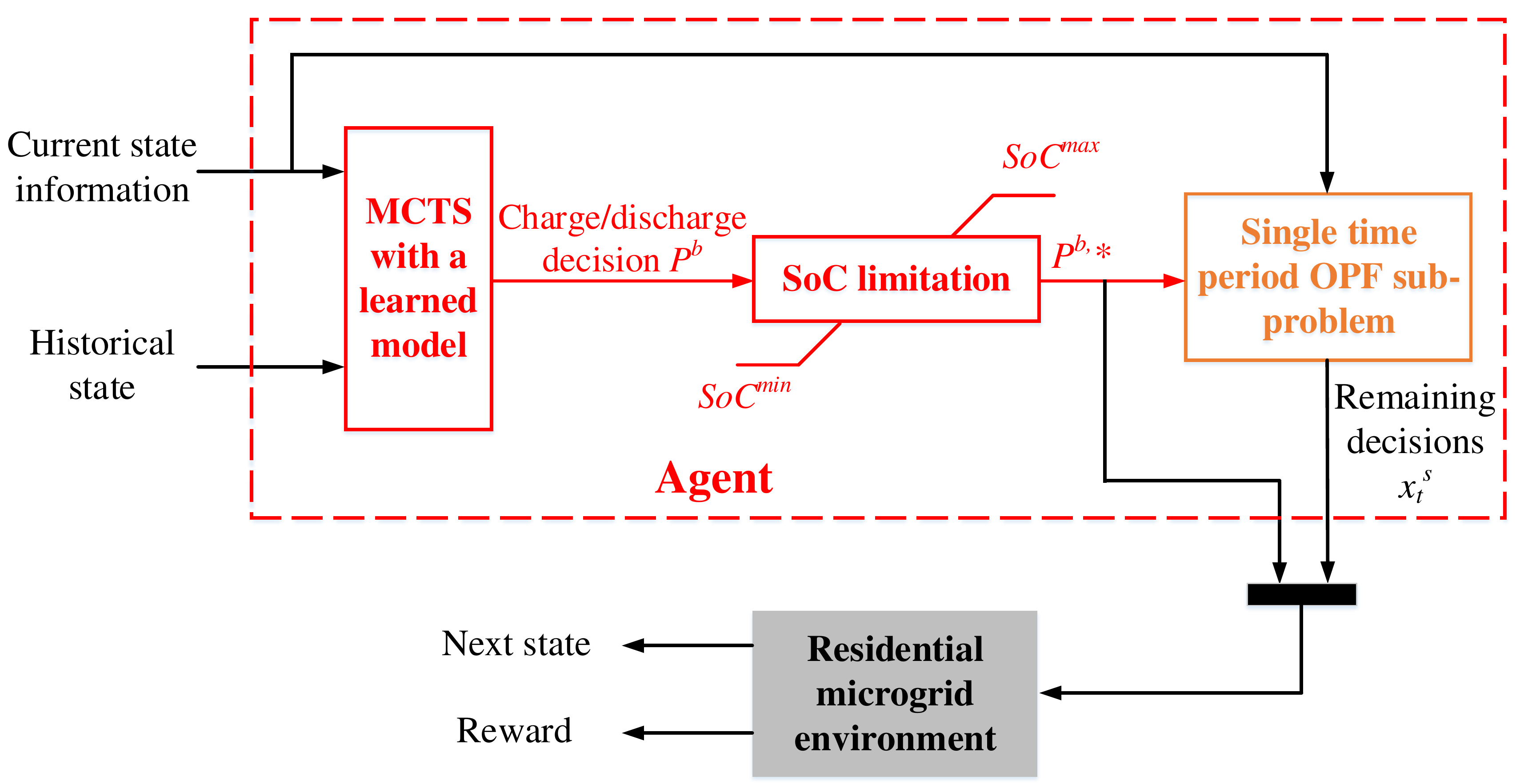}
\caption{The online optimization process at each time-step.} \label{fig:MCTS_OPF}
\vspace{-0.8em}
\end{figure}

To solve the challenges, we propose to divide the decision variables shown in (\ref{EQ26}) into two categories.
The first category contains the active charge/discharge decision of the battery $P^b(t)$, which is determined directly by the model based reinforcement learning algorithm.
And we discretize the charge/discharge decision to facilitate the application of MCTS algorithm.
Researchers have proposed a MCTS algorithm that can deal with continuous actions \cite{rajamaki2018continuous}.
The extension, however, is left for future work.
The second category includes the remaining decisions in (\ref{EQ26}), which are optimized as a single time period optimal power flow (OPF) sub-problem, as shown in (\ref{EQ32}).
\begin{equation}\label{EQ32}
\begin{aligned}
\min_{x_t^s} \quad & \bigg( \sum_{g \in \mathcal{G}} C_g^{DG}\big(P_g^{DG}(t)\big) + C_{grid}\big(P_{buy}^{grid}(t), P_{sell}^{grid}(t)\big) \\ 
   &+ C_{bat}\big(P^{b,*}(t)\big) + C_{cur}\big( P_{cur}^{ren}(t)\big) \bigg)
   \\
    \mbox{s.t.}\quad & (\ref{EQ6})-(\ref{EQ13}), (\ref{EQ17}), (\ref{EQ20})-(\ref{EQ22}), \emph{and} \, (\ref{EQ24})
\end{aligned}
\end{equation}
where, $x_t^s$ denotes the decisions optimized in the OPF sub-problem.
$P^{b,*}(t)$ is the optimal decision directly given by the MB-DRL method.
The sub-problem (\ref{EQ32}) can be solve by second-order cone programming (SOCP) technique.
Note that the active charge/discharge decision is fixed when solving the OPF sub-problem.
The proposed optimization framework that combines the MCTS with the mathematical optimization technique will bring two advantages.
The first advantage is that the action space handled by the MCTS is largely reduced.
The second one is that most constraints can be handled by OPF sub-problem, which avoids letting the MCTS and the neural network model directly deal with a large number of constraints and this will greatly help the learning process of the agent.
The schematic diagram of the proposed MB-DRL based online optimization algorithm for RM is shown in Fig. \ref{fig:MCTS_OPF}.
In the figure, the charge/discharge decision $P^b(t)$ given by the MCTS algorithm is checked by the overcharge/overdischarge limitation unit to ensure the constraint (\ref{EQ19}) is fulfilled, as shown in (\ref{EQ33}).
\begin{equation}\label{EQ33}
\begin {aligned}
   P^{b,*} (t) = \left\{ 
   \begin{array}{lr} 
   P^{b}(t), \ SoC^{min} \le SoC(t+\Delta t) \le SoC^{max} \\
   -\frac{E^{max}\cdot(SoC^{max}-SoC(t))}{\eta^{ch} \Delta t}, \ SoC(t+\Delta t) > SoC^{max} \\
   \frac{(SoC(t)-SoC^{min})\cdot E^{max} \cdot \eta^{dis}}{\Delta t}, \ SoC(t+\Delta t) < SoC^{min}
   \end{array}
\right.
\end {aligned}
\end{equation}
where, $SoC(t+\Delta t)$ can be computed using (\ref{EQ18}) according to $P^{b}(t)$.

The multi-time period optimization problem shown in (\ref{EQ25}) is decomposed to multiple single time period sub-problems in this work according to Bellman's optimality principle, and the optimal decisions of each time-step are optimized using MCTS planning method combined with the learned neural network model, as shown in \textbf{Algorithm 1}.
We assume all the states of the RM are observable.
At time-step $t$, the states of the RM are collected, and are fed to the representation network to compute the internal state of the system using (\ref{EQ29}).
Then, the root node is created according to the obtained internal state $\hat s_t^0$.
Starting from the root node, a search tree is constructed by iteratively performing $N$ simulations (see lines 6 to 11 of \textbf{Algorithm 1}).
And the optimal output power of the battery can be determined according to the visit frequency of the child nodes of the root node.
Then use (\ref{EQ33}) to make sure the SoC limitation will not be violated.
Finally, the remaining decisions $x_t^s$ are determined by solving the OPF sub-problem, and the optimal decisions are executed.
In the next time-step, the above optimization process is repeated.

We can find that the key step is the search tree building process (see lines 4 to 11 of \textbf{Algorithm 1}).
The tree consists of nodes and edges.
Each node is associated with an internal state $\hat s$.
For each decision $\hat x$ from $\hat s$ there is an edge $(\hat s, \hat x)$ that leads to a child node.
The information stored in the edge $(\hat s, \hat x)$ is $\omega = \{N(\hat s, \hat x), Q(\hat s, \hat x), P(\hat s, \hat x), R(\hat s, \hat x), S(\hat s, \hat x)\}$.
$N$, $Q$, $P$, $R$, and $S$ represent visit counts, mean value, policy, reward, and state transition, respectively.
For simplicity, $\hat s_t^k$ and $\hat x_t^k$ will be represented by $\hat s^k$ and $\hat x^k$ in the following context.
Each simulation includes the following three steps: 

\begin{breakablealgorithm}   \caption{The Proposed MB-DRL based Online Optimization Algorithm.}\label{alg:MicrogridZero}
\small
\begin{algorithmic}[1]
    \State Load the pre-trained neural network model which includes the representation network, the dynamic network, and the prediction network; load the RM system parameters.
	\For {$t = {\Delta t, 2 \Delta t, \cdots, T}$}:
			\State Get the current state information of the system. \Comment {\textit{{\footnotesize (\ref{EQ27}).}}}
			\State {Compute the internal state of the RM according to the current
			\Statex \quad \ \ state information and the historical state information. \Comment {\textit{{\footnotesize (\ref{EQ29}).}}}}
			\State {Create the root node according to the computed internal state, 
			\Statex \quad \ \ and set $n=1$.}
			\While {$n \leq N$}
				\State {Perform the selection step start from root node. \Comment {\textit{{\footnotesize (\ref{EQ34}).}}}}
				\State Perform the expansion step. \Comment {\textit{{\footnotesize (\ref{EQ35})-(\ref{EQ37}).}}}
				\State Perform the backpropagation step. \Comment {\textit{{\footnotesize (\ref{EQ38})-(\ref{EQ40}).}}}
				\State $n = n + 1$.
			\EndWhile
			\State {Get the optimal charge/discharge decision $P^{b,*}(t)$ that corre- 
			\Statex \quad \ \ spond to the most visited child nodes of the root node.}
			\State {Overcharge/overdischarge check and get the optimal decision 
			\Statex \quad \ \ $P^{b,*}(t)$. \Comment {\textit{{\footnotesize (\ref{EQ33}).}}}}
			\State {Fix the charge/discharge decision of the battery to be $P^{b,*}(t)$ 
			\Statex \quad \ \ and solve the OPF sub-problem (\ref{EQ32}) to compute the optimal 
			\Statex \quad \ \ decisions $x_t^{s,*}$.}
			\State {Execute the optimal decisions and calculate the next state
			\Statex \quad \ \ of the RM system.}
	\EndFor 
\end{algorithmic}
\end{breakablealgorithm}

\textbf{1) Selection}: Starting from the root state $\hat s^0$, the simulation selects a decision and reaches to the corresponding child node. The Selection step finishes when the simulation arrives at a leaf node $\hat s^l$.
For each hypothetical time-step $k = 1, 2, \cdots, l$ of the simulation, a decision $\hat x^k$ is selected using (\ref{EQ34}).
\begin{equation}\label{EQ34}
\begin {aligned}
   \hat x^k = \arg \max_{x} [Q(\hat s, \hat x) + P(\hat s, \hat x) \cdot \frac{\sqrt{\sum_{\varsigma} N(\hat s, \varsigma)}}{1+N(\hat s, \hat x)} \cdot \\ (c_1 + log(\frac{\sqrt{\sum_{\varsigma} N(\hat s, \varsigma) + c_2 + 1}}{c_2}))]
\end {aligned}
\end{equation}
where, $Q(\hat s, \hat x)$ is the value that takes decision $\hat x$ from state $\hat s$, which represents the average reward for taking this decision (exploitation term).
The second part of the right-hand side of (\ref{EQ34}) is the exploration term which can encourage the simulation to take decisions that have been less selected.
$P(\hat s, \hat x)$ is the prior probability of taking decision $\hat x$ from state $\hat s$.
$c_1$ and $c_2$ are the constant value which is used to balance the exploitation and exploration term.
In general, we can set $c_1 = 1.25$ and $c_2 = 19652$, as suggested by reference \cite{schrittwieser2019mastering}.
For $k<l$, the next state and reward are obtained by looking up the state transition table $\hat s^k = S(\hat s ^{k-1}, \hat x^k)$ and corresponding reward table $r^k = R(\hat s^{k-1}, \hat x^k)$, respectively.

\textbf{2) Expansion}: When the simulation steps to the leaf node of the tree at the final time-step $l$, a new child node will be added to the tree. The state of the new node $\hat s^l$ is computed by the learned dynamic network as shown in (\ref{EQ35}). 
Besides, the corresponding reward $\hat r^l$ is also computed, and the obtained new internal state and reward will be stored in the corresponding tables, $R(\hat s^{l-1}, \hat x^l) = \hat r^l$, $S(\hat s^{l-1}, \hat x^l) = \hat s^l$.
\begin{equation}\label{EQ35}
\begin {aligned}
   \hat r^l, \hat s^l = g_{\vartheta} (\hat s^{l-1}, \hat x^l)
\end {aligned}
\end{equation}
The policy and value correspond to state $\hat s^l$ are computed using the learned prediction network:
\begin{equation}\label{EQ36}
\begin {aligned}
   \textbf{p}^l, v^l = f_{\theta}(\hat s^l)
\end {aligned}
\end{equation}
Then, the information stored in the new edge $(\hat s^l, \hat x)$ is initialized to:
\begin{equation}\label{EQ37}
	\omega = \{ N(\hat s^l, \hat x) = 0, Q(\hat s^l, \hat x) = 0, P(\hat s^l, \hat x) = \textbf{p}^l \}.
\end{equation}

\textbf{3) Backpropagation}: At the end of the simulation, the information of the new leaf node is backpropagated along the trajectory to update the statistics of all the edges in the simulation path.
For $k=l, \cdots, 1$, the statistics for each edge $(\hat s^{k-1}, \hat x^k)$ along the trajectory are updated by:
\begin{equation}\label{EQ38}
\begin {aligned}
	Q(\hat s^{k-1}, \hat x^k) &= \frac{N(\hat s^{k-1}, \hat x^k) \cdot Q(\hat s^{k-1}, \hat x^k) + G^k}{N(\hat s^{k-1}, \hat x^k) + 1} \\
N(\hat s^{k-1}, \hat x^k) &= N(\hat s^{k-1}, \hat x^k) + 1	
\end {aligned}
\end{equation}
where, $G^k$ represents the $l-k$-step estimation of the cumulative discounted reward, bootstrapping from the value $v^l$,
\begin{equation}\label{EQ39}
	G^k = \sum_{\tau = 0}^{l-1-k} \gamma^{\tau} \hat r^{k+1+\tau} + \gamma^{l-k} v^l	
\end{equation}
In (\ref{EQ39}), $\gamma$ is the discount factor and $0 < \gamma < 1$.
For the online optimization problem of this paper, the value functions $Q(\hat s, \hat x)$ is unbounded, which makes the pUCT rule shown in (\ref{EQ34}) cannot perform properly.
To avoid this, a normalized value $\bar Q \in [0,1]$ is adopted in the pUCT rule.
$\bar Q$ is computed by using the minimum-maximum values observed in the search tree up to that point:
\begin{equation}\label{EQ40}
	\bar Q(\hat s^{k-1}, \hat x^k) = \frac{Q(\hat s^{k-1}, \hat x^k) - \min_{\hat s, \hat x \in{Tree}} Q(\hat s, \hat x)}{\max_{\hat s, \hat x \in{Tree}} Q(\hat s, \hat x) - \min_{\hat s, \hat x \in{Tree}} Q(\hat s, \hat x)}	
\end{equation}

In general, the proposed RM online optimization algorithm utilizes the MCTS algorithm and a learned model to search over hypothetical future trajectories $\hat x^1, \cdots, \hat x^k$ given historical state observations $s_1, \cdots, s_t$, and outputs a recommended policy $\pi_t$ and value estimation $v_t$.
Then, a charge/discharge decision is sampled from the policy $\pi_t$, and we use SOCP technique to obtain the remaining decisions.
The process is repeated till the end of the optimization horizon.
Inside the search tree, the representation network is used to generate an initial internal state in order to improve the prediction network performance.
The policy and value estimates computed by prediction network are used by each internal node to select and build its child nodes, and the dynamic network is adopted to compute the next state $\hat s$ and reward $\hat r$ after taking a decision $\hat x$.
It can be found that, with the help of the learned model, the agent does not use forecast information of renewable energy generation and load power during the decision making process, and can make decisions without prior knowledge of the uncertainties in the microgrid system.

\subsection{Designation of the Neural Network Model}
From \textbf{Algorithm 1}, one can find that the learned model is critical for the proposed algorithm.
To obtain a good model, two things are very important.
The first one is to design a suitable network architecture for the representation network, dynamic network, and prediction network.
The second one is the off-line training of the designed model.
In this subsection, we will design an appropriate neural network model architecture for the optimization problem in this work, while the model training method will be presented in the next subsection.

\begin{figure}[t]\centering
\includegraphics[width=3.5in]{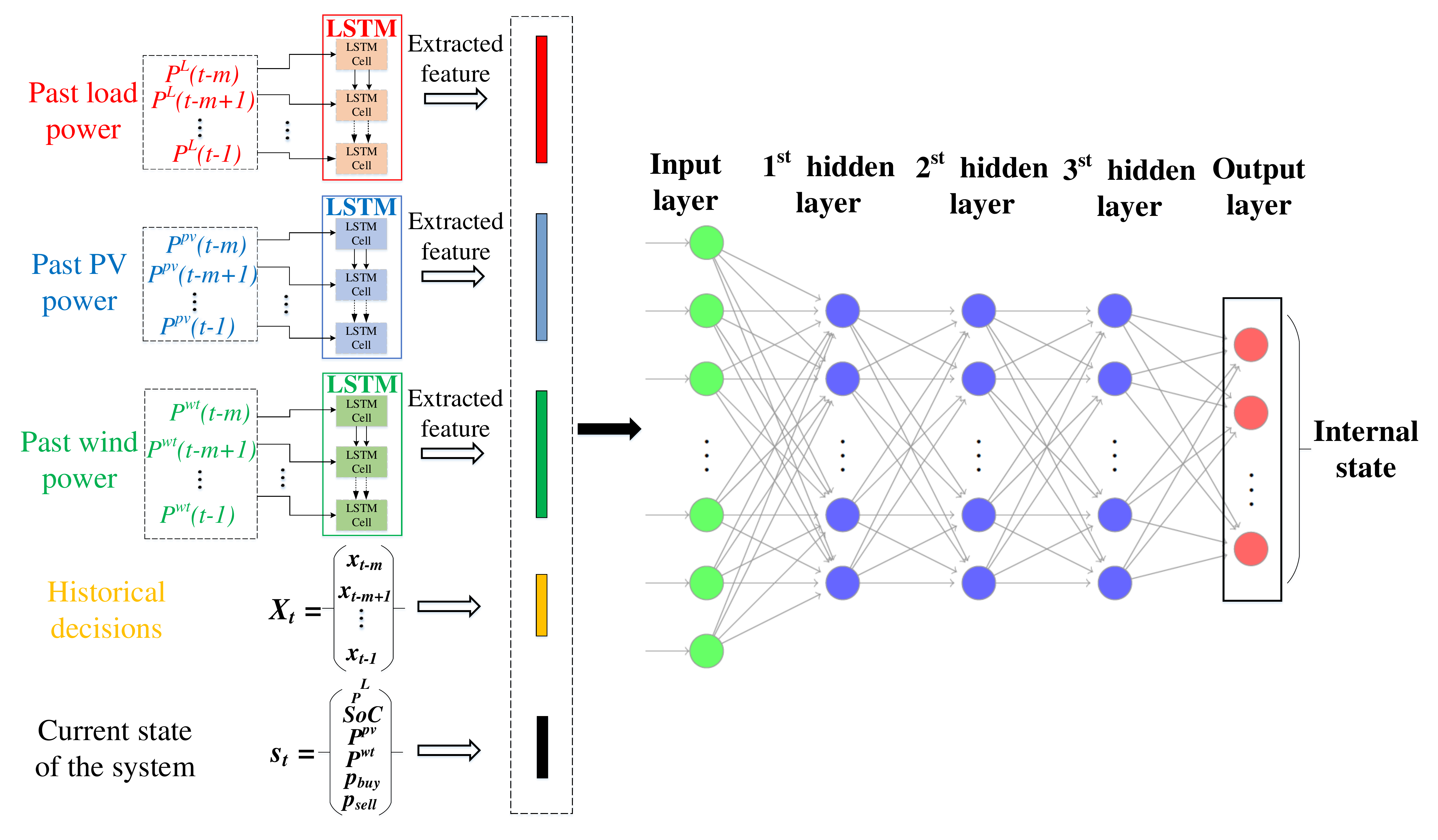}
\caption{The architecture of the designed representation network.} \label{fig:RepresentationNetwork}
\vspace{-0.5em}
\end{figure}

The internal state is computed using the representation network, which will affect the prediction accuracy of the future quantities (including policies, values, and rewards).
So, the representation network needs to be carefully designed.
In \cite{schrittwieser2019mastering}, the convolution neural network and residual blocks are adopted to play Atari games since the observations have large spatial resolutions and a strong spatial correlation.
However, different from Atari games, the observations of the RM system have strong temporal correlation.
For instance, we find strong correlations between wind/solar power in adjacent time periods.
Considering this, we propose to use Long Short-Term Memory (LSTM) networks to extract features from the historical PV generation power, wind power, and load power.
Then the current system state shown in (\ref{EQ1}) is concatenated with the extracted features and will be fed to a multi-layer neural network.
The layout of the representation network is shown in Fig. \ref{fig:RepresentationNetwork}.
The output of the representation network is an $O_{rep}$-dimensional vector.
As we mentioned in Section III(B), the internal state output by the representation network does not have any physical meaning.
Unlike using the LSTM networks to directly forecast the PV/wind/load power in \cite{SuLSTM2019}, there is no future information of PV/wind/load power is used as the label data during the training process.
Hence, this procedure conducted by the representation network is not forecasting future power and it can be just regarded as a feature extraction from historical and current system state.

\begin{figure}[t]\centering
\includegraphics[width=3.5in]{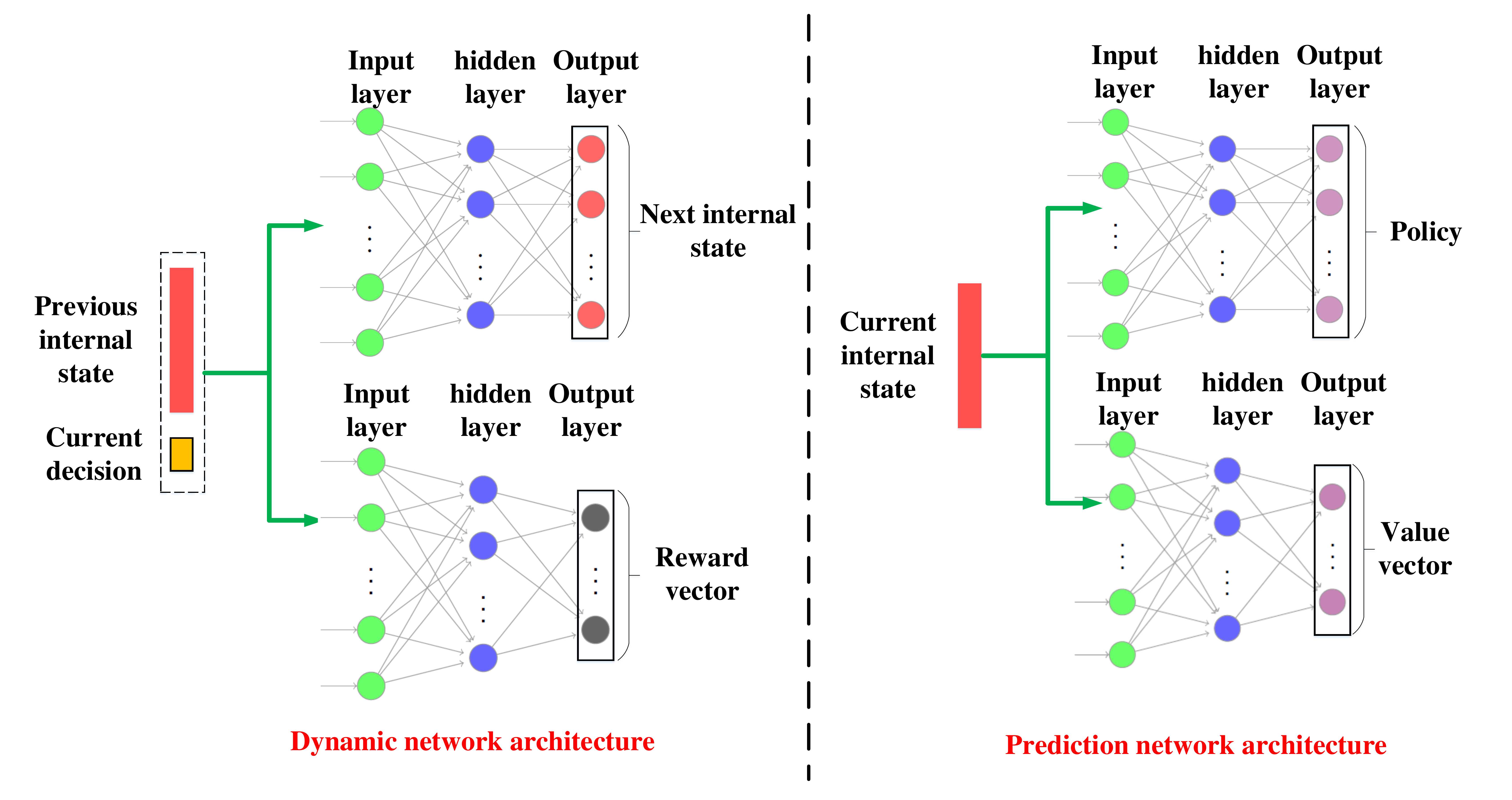}
\caption{The architecture of the designed prediction network and the dynamic network.} \label{fig:PredictionNetwork}
\end{figure}

For the dynamic network shown in Fig. \ref{fig:PredictionNetwork}, the current decision stacked with the internal state of the previous hypothetical step are set as input, and two individual multi-layer neural networks are adopted to respectively compute the next internal state and the reward.
The output of the reward computing network is an $O_r$-dimensional vector.
The dynamic network can be regarded as an approximator of the RM system, which enables the proposed model-based learning algorithm to make optimal decisions without knowledge of the system dynamics.

The prediction network shown in Fig. \ref{fig:PredictionNetwork} also contains two individual multi-layer neural networks.
One is used to compute the probability of each charge/discharge decision being selected, and the other is to get the value $v^k$ in (\ref{EQ31}).
The output of the network can estimate the possible total rewards after current time-step and predict which of the currently available charge/discharge decisions are likely to work best.
The output of the policy computing network and value computing network is an $O_{p}$-dimensional vector and an $O_{v}$-dimensional vector, respectively.
Note that the prediction network (as named in \cite{schrittwieser2019mastering}) here is used to compute the policy $\textbf{p}$ and value $v$ to help the agent makes optimal decisions, unlike the neural network based renewable power forecasting module predicting the future wind/solar power. 
Finally, the designed representation network, dynamic network, and prediction network formed the model that utilized in MCTS.
\subsection{Training Method of the Proposed MB-DRL Algorithm}
To obtain a good online optimization performance, the model needs to be well-trained off-line first.
The model is trained by reinforcement learning from data of self-play.
The parameters of the representation network, dynamic network, and prediction network are trained jointly to accurately match the computed policy, value, and reward, for every hypothetical step $k$, with corresponding target values observed when $k$ actual time-steps have elapsed.
More specifically, the training objective is to minimize the following errors by updating network weights:
\begin{equation}\label{EQ41}
\begin {aligned}
	l_t(\theta, \vartheta, \phi) = \sum_{k=0}^K [ l^r (r_{t+k}, \hat r_t^k) + l^v (z_{t+k}, v_t^k) + l^p (\pi_{t+k}, p_t^k) \\ 
	+ c (\left \| \theta \right \|^ 2 + \left \| \vartheta \right \|^ 2 + \left \| \phi \right \|^ 2 ) ]
\end {aligned}
\end{equation}
where, $r_{t+k}$ represents the improved reward target, that is, the observed reward after an actual time-step.
$\hat r_t^k$ is the predicted reward that computed by dynamic network. 
$z_{t+k}$ represents the improved value target that can be computed by adding up $n$ step discounted rewards and the corresponding search value, $z_t = r_{t+1} + \gamma r_{t+2} + \cdots + \gamma^{n-1} r_{t+n} + \gamma^{n} v_{t+n}$.
$\pi_{t+k}$ represents the improved policy target that is generated by an MCTS search.
$v_t^k$, $p_t^k$ are the predicted value and policy that computed by prediction networks, respectively.
So, the first three parts of the right-hand side of the equation (\ref{EQ41}) represents the errors between the predicted reward/value/policy and the reward/value/policy target.
The errors can be computed as (\ref{EQ42}).
The last part of (\ref{EQ41}) represents the L2 regulation term of the network weights.

\begin{equation}\label{EQ42}
\left\{ 
   \begin{array}{lr}
	l^r(r,\mathbf{\hat r}) = \varphi (\varrho(r))^T \log \mathbf{\hat r} \\
	l^v(z,\textbf{q}) = \varphi (\varrho(z))^T \log \textbf{q} \\
	l^p(\pi,\textbf{p}) = \pi^T \log \textbf{p}
	\end {array}
\right.
\end{equation}
where, $\mathbf{\hat r}$ is the reward vector computed by the dynamic network;
$\textbf{p}$ and \textbf{q} are the policy and value vector computed by the prediction network.
$\varrho(r)$ is an invertible function that used to scale the number $r$, where $\varrho(r) = sign(r)(\sqrt{\left|r\right|+1}-1+\epsilon r)$ with $\epsilon = 10^{-3}$.
The function $\varrho(\cdot)$ is introduced to reduce the variance of the optimization target, which can help to improve algorithm convergence \cite{pohlen2018observe}.
$\varphi (\cdot)$ refers to the transformation function that used to transfer the scalar reward and value targets to equivalent categorical representations.
If we use a discrete support set of size $2n+1$ with one support for every integer between -$n$ and $n$, by applying $\varphi (b)$, a real number $b$ ($-n \leq b \leq n$) can be represented through a linear combination of its two adjacent integers $\lfloor b \rfloor$ and $\lceil b \rceil$, so that the original scalar can be recovered by $b = \lfloor b \rfloor \cdot (\lceil b \rceil - b) + \lceil b \rceil \cdot (b - \lfloor b \rfloor)$.
Besides, to maintain similar gradient magnitude across different unroll steps, we adopted the gradient scaling method proposed in Appendix G of reference \cite{schrittwieser2019mastering}.   

\begin{figure}[t]\centering
\includegraphics[width=3.2in]{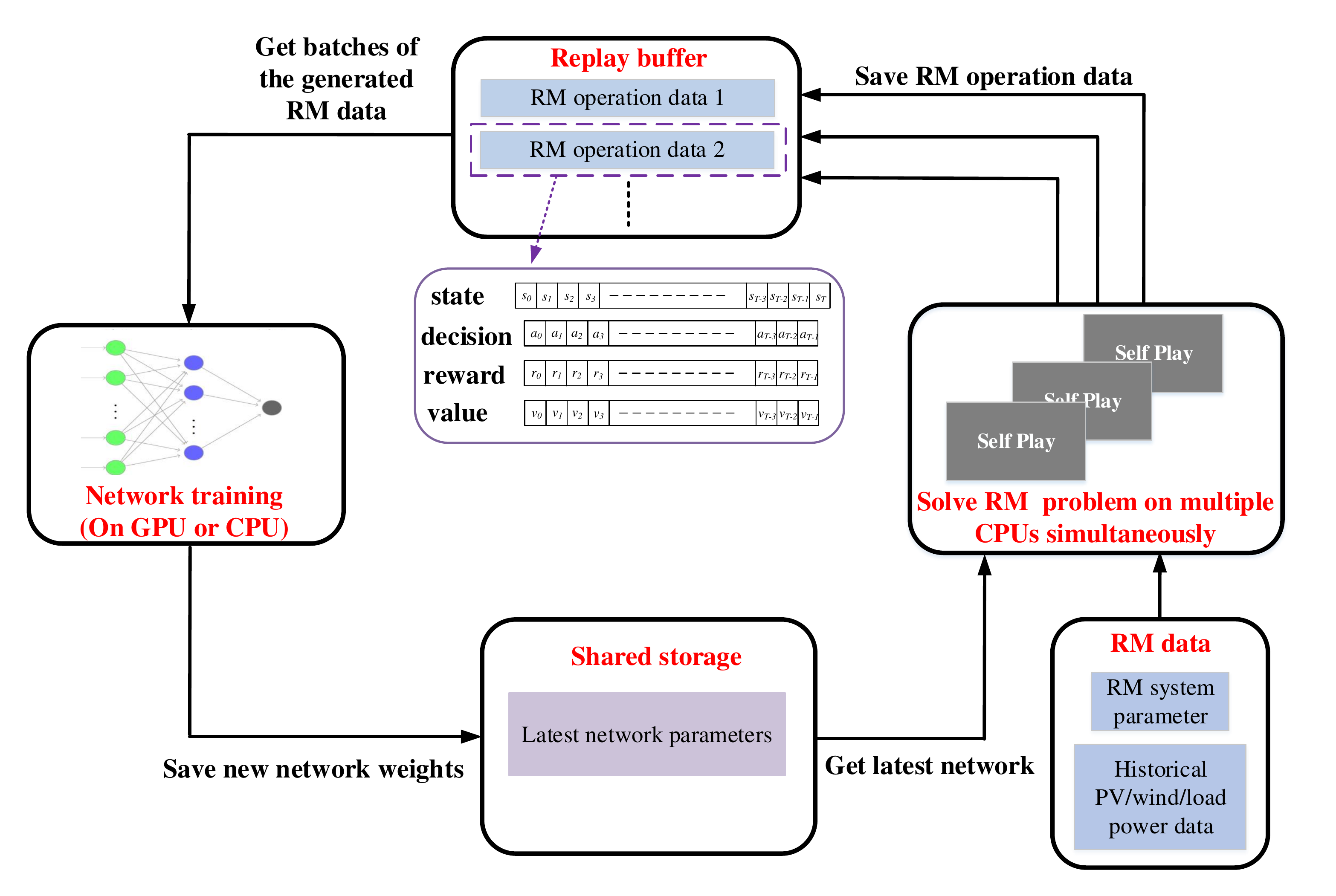}
\caption{The training architecture of the model.} \label{fig:TrainingArch}
\end{figure}

The model training process is shown in Fig. \ref{fig:TrainingArch}.
The training can be split into two independent parts: network training (producing an improved neural network model) and self-play (generating RM operation data).
The generated self-play data is stored in the replay buffer, and we sample training data from the buffer to update the neural network model (see Eq. (\ref{EQ41}) - (\ref{EQ42})).
Then, the updated neural network is stored in the shared storage, and a random microgrid scenario will be selected by the self-play units to generate new RM operation data with the latest network.
The process repeated until the algorithm converged.
To speed up the training, network training and self-play can be performed in parallel.
The pseudocode of the training method of the proposed algorithm has been given in \textbf{Algorithm 2} - \textbf{Algorithm 4} (see Appendix).
 
\section{Numerical Analysis}  \label{Numerical Analysis}
\begin{figure}[t]\centering
\includegraphics[width=3.5in]{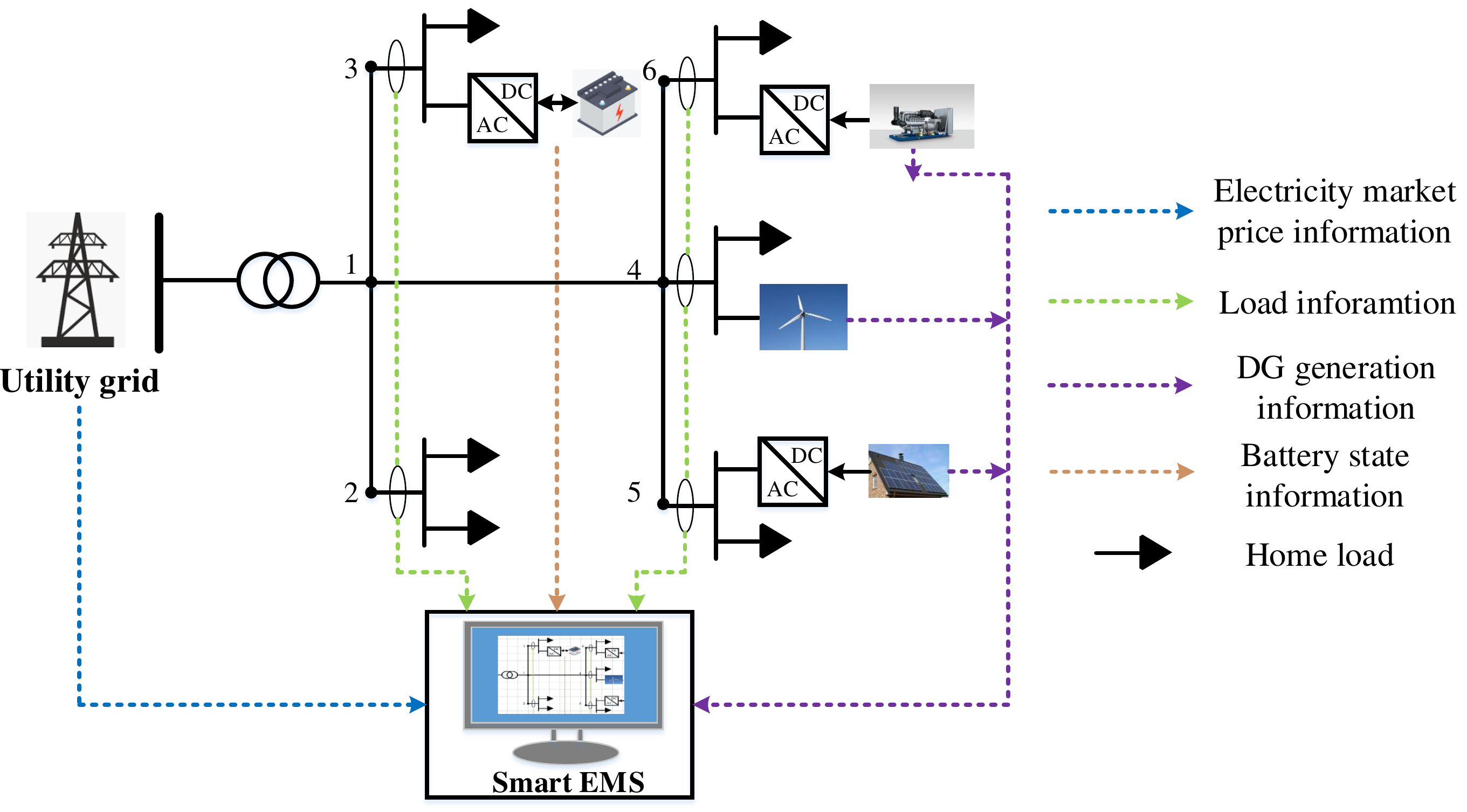}
\caption{The residential microgrid system.} \label{fig:microgridSch}
\end{figure}

\begin{figure}[t]\centering
\includegraphics[width=3.5in]{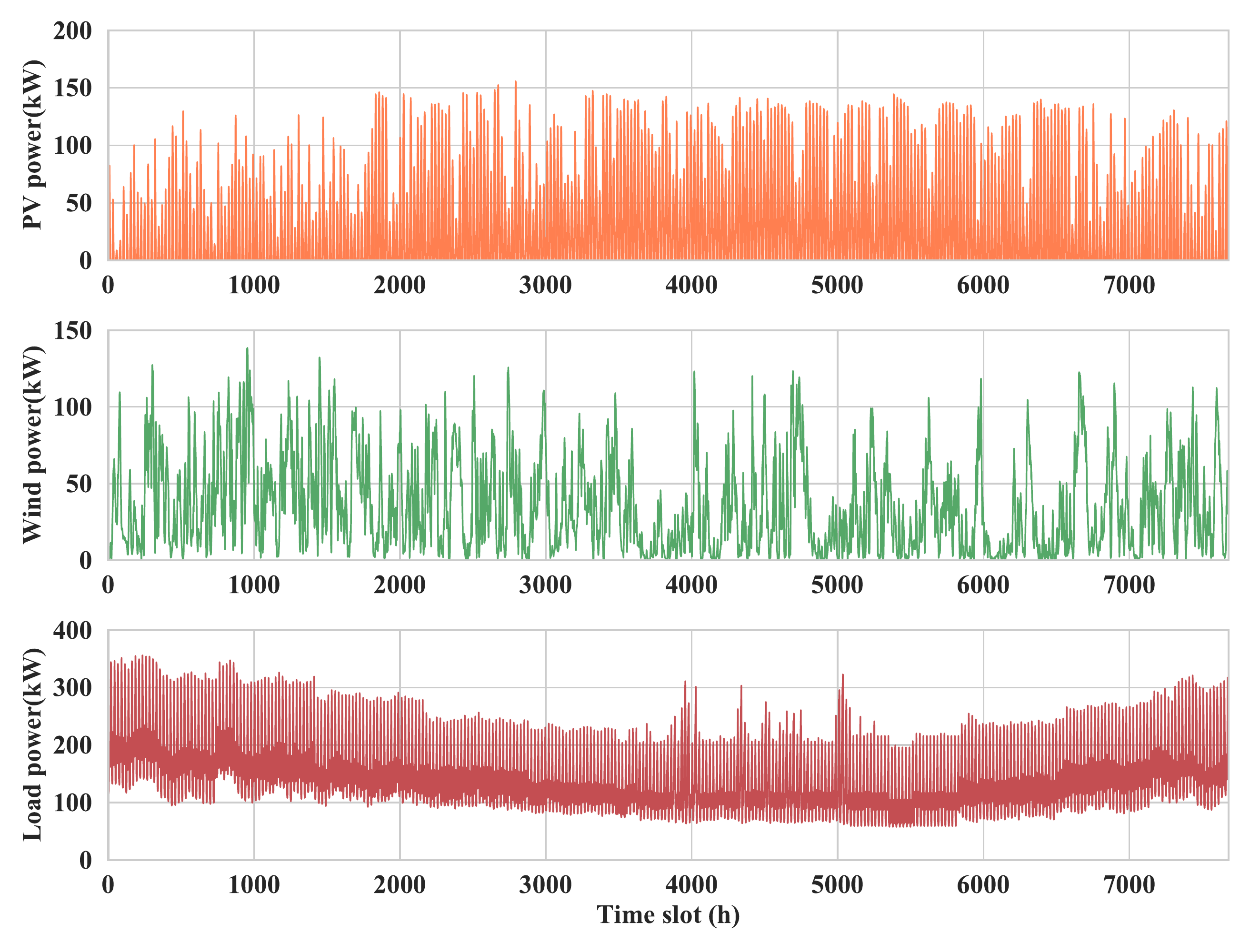}
\caption{The training data profiles of solar power, wind power, and load power. The historical wind and PV power data \cite{WTPV_Data} and the residential load data \cite{LoadTraining} from Jan. 1, 2016 to Nov. 15, 2016 are used as training data set.} \label{fig:TrainingDataProfile}
\end{figure}

\begin{figure}[t]\centering
\includegraphics[width=3.5in]{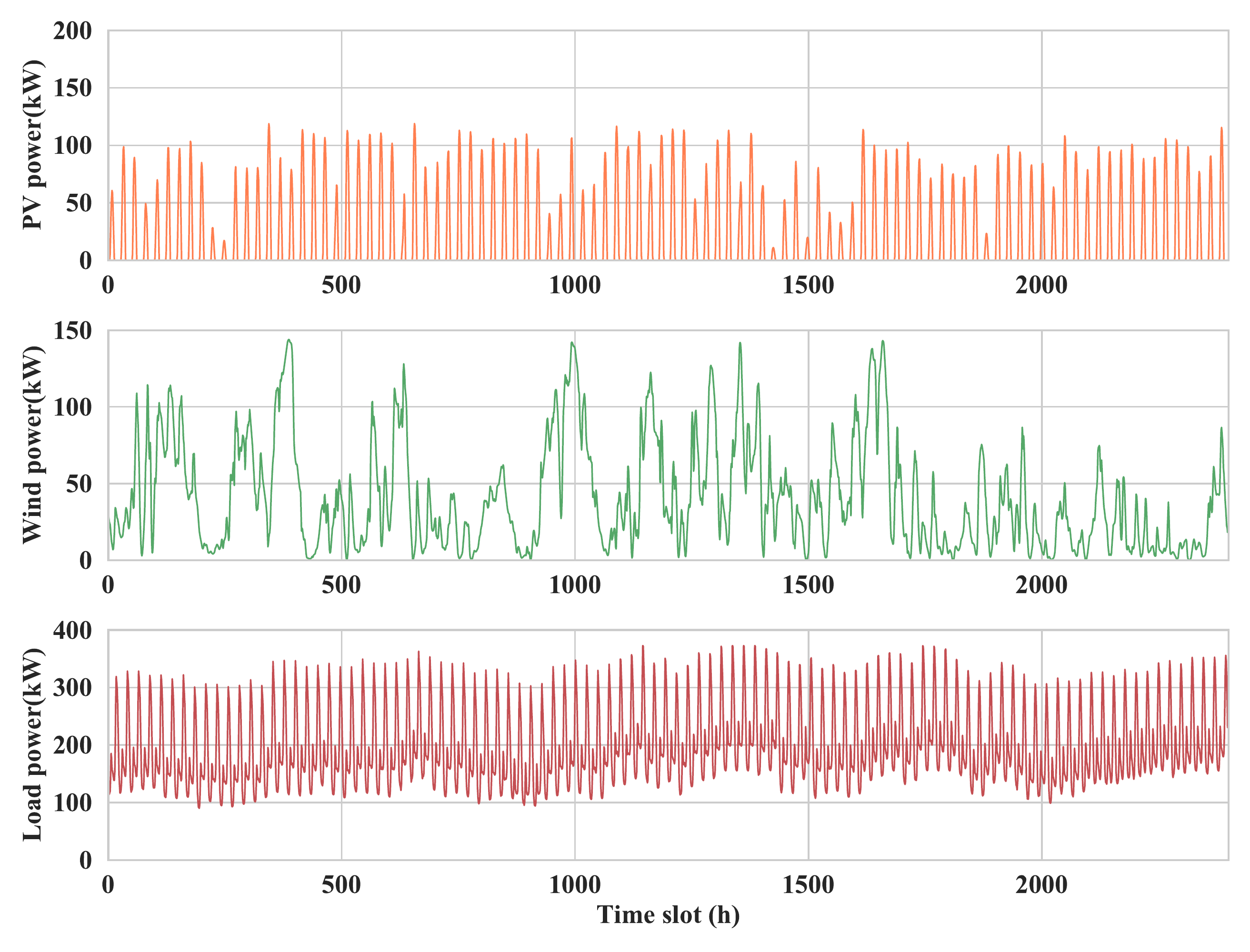}
\caption{The testing data profiles of solar power, wind power, and load power. The historical wind and PV power data \cite{WTPV_Data} from Nov. 7, 2015 to Dec. 31, 2015 and Nov. 16, 2016 to Dec. 31, 2016 are used as testing data set. The historical load power data \cite{LoadTraining} from Nov. 16 to Dec. 31, and the historical load power data \cite{LoadTesting} from Jan. 1 to Feb. 24 are used as testing data set.
} \label{fig:TestingDataProfile}
\end{figure}

In this section, the online optimization performance of the proposed algorithm is validated on an RM test system shown in Fig. \ref{fig:microgridSch}.
In the RM system, the electricity is provided by the distributed renewable energy, diesel generator, energy storage system (ESS), and the utility grid.
The diesel generator is connected to bus 6, and its upper and lower generation power are 30 kW and 10 kW, respectively.
The fuel cost coefficients of the diesel generator are $\alpha_g = 1.04 \ \$/kW^2 h$, $\beta_g = 0.03 \ \$/kWh$, $c_g = 1.3 \ \$/h$.
The ESS is a 500 kWh @100kW battery with a round-trip efficiency of 90.25\%, which is connected to bus 3.
To ensure the life span of the battery, the minimum stored energy in the battery is set to be 100 kWh.
The degradation cost coefficient is set to be 0.1 \$/kWh \cite{Du8769895}.
The charging/discharging power of the battery is divided into 9 levels (-100, -75, -50, -25, 0, 25, 50, 75, 100) $kW$.
The negative value represents the charging mode.
The load data used in the simulation are the historical residential load profiles of Anchorage Alaska State from \cite{LoadTraining,LoadTesting}.
To simplify the simulation, we assume that the power demand of each bus accounts for a fixed proportion of the total demand, and the power factor of each load is constant.
The active power demand ratio of bus 2 to 6 are 20\%, 10\%, 30\%, 20\% and 20\%, respectively.
Solar power and wind power data are actual historical data from \cite{WTPV_Data}.
The profiles of the data are shown in Fig. \ref{fig:TrainingDataProfile} and Fig. \ref{fig:TestingDataProfile}.
The dynamic market energy price of Southern California residential area \cite{Price} is adopted, and the selling price is set to be 50\% of the market price, as shown in Table \ref{price}. 
The power network resistance and reactance parameters are given in Table \ref{Resistance}.  

\begin{table}\centering
\footnotesize
\newcommand{\tabincell}[2]{\begin{tabular}{@{}#1@{}}#2\end{tabular}}
\caption{The market energy price and the selling price ($\$/kWh$). }\label{price}
\setlength{\tabcolsep}{0.5mm}{
\begin{tabular}
{c|c|c|c|c} \hline \hline
\tabincell{c}{$Time$ $periods$ } &8:00 - 14:00 &14:00 - 20:00 &20:00 - 22:00 &22:00 - 8:00 \\ \hline	
\tabincell{c}{$Marke$t $energy$ \\ $price$ } &0.28 &0.48 &0.28 &0.12 \\ \hline	
\tabincell{c}{$Selling$ $price$}  & 0.14    &  0.24  &  0.14  &  0.06  \\ \hline \hline
\end{tabular}}
\end{table}

\begin{table}\centering
\footnotesize
\newcommand{\tabincell}[2]{\begin{tabular}{@{}#1@{}}#2\end{tabular}}
\caption{The network resistance and reactance parameters.}\label{Resistance}
\setlength{\tabcolsep}{0.5mm}{
\begin{tabular}
{c|c|c|c|c} \hline \hline
\tabincell{c}{Cable} &\tabincell{c}{From bus} &\tabincell{c}{To bus} &\tabincell{c}{Resistance ($10^{-2}\Omega$)} &\tabincell{c}{Reactance ($10^{-2}\Omega$)} \\ \hline	
\tabincell{c}{$L_1$ } &1 &2 &0.922 &0.470 \\ \hline	
\tabincell{c}{$L_2$}  & 1    &  3  &  4.930  &  2.511  \\ \hline
\tabincell{c}{$L_3$}  & 1    &  4  &  3.660  &  1.864  \\ \hline
\tabincell{c}{$L_4$}  & 4    &  5  &  3.811  &  1.941  \\ \hline
\tabincell{c}{$L_5$}  & 4    &  6  &  1.872  &  6.188  \\ \hline \hline
\end{tabular}}
\end{table}



In this work, the encoding size of the internal state is set to 10, which means the internal state computed by representation network and dynamic network is a 10-dimensional vector.
The support sizes of the reward and value that computed by the prediction network are also set to 10 in this simulation.
Three independent LSTM units are adopted in the representation network to extract features from the past 6-h PV power, wind power, and active load power, respectively.
The network parameters adopted in the simulation are shown in Table \ref{MicrogridZero-Model}.
The hyperparameters of the algorithm are set as follows: batch size $B=64$, learning rate $\alpha=0.005$, discount factor $\gamma=0.997$, number of simulations per search $N=20$, number of unrolled hypothetical steps $K=5$, bootstrapping steps $n=10$, ratio of self-play speed to training speed is 0.1.
All the simulations are conducted on an Intel Core i7-4790 @ 3.60 Ghz $\times$ 8 Ubuntu based minitower computer with 32 GB RAM.
For the online optimization problem, we use 1 CPU for training and 7 CPUs for self-play.
The code is written in Python with PyTorch.
\begin{table}[t]\centering
\vspace{-0.8em}
\caption{The network parameters of the proposed algorithm.} \label{MicrogridZero-Model}
\includegraphics[width=3.0in]{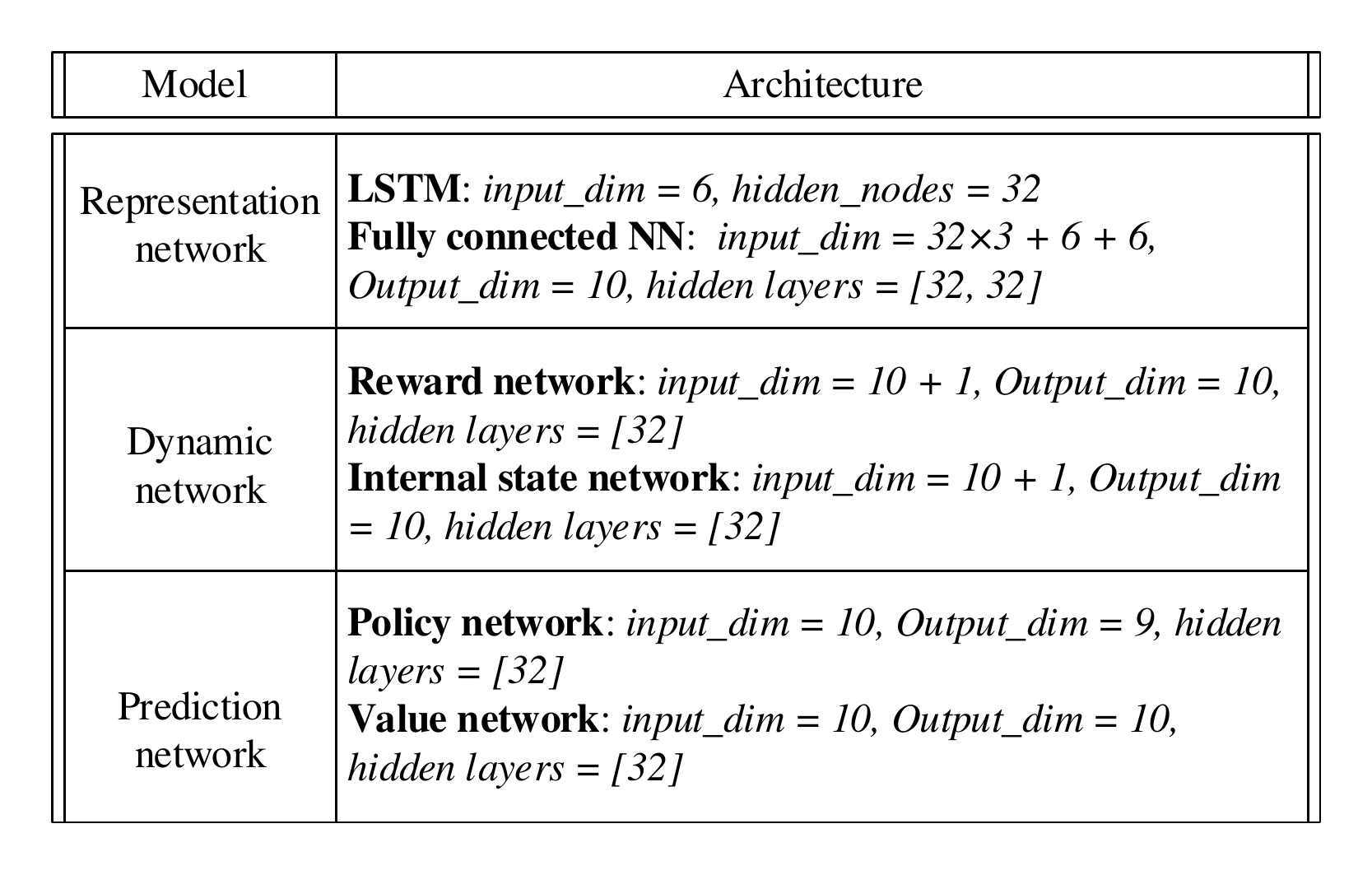}
\vspace{-1em}
\end{table}

\subsection{Training Result of the Proposed Algorithm}
\begin{figure}[t]\centering
\includegraphics[width=3.5in]{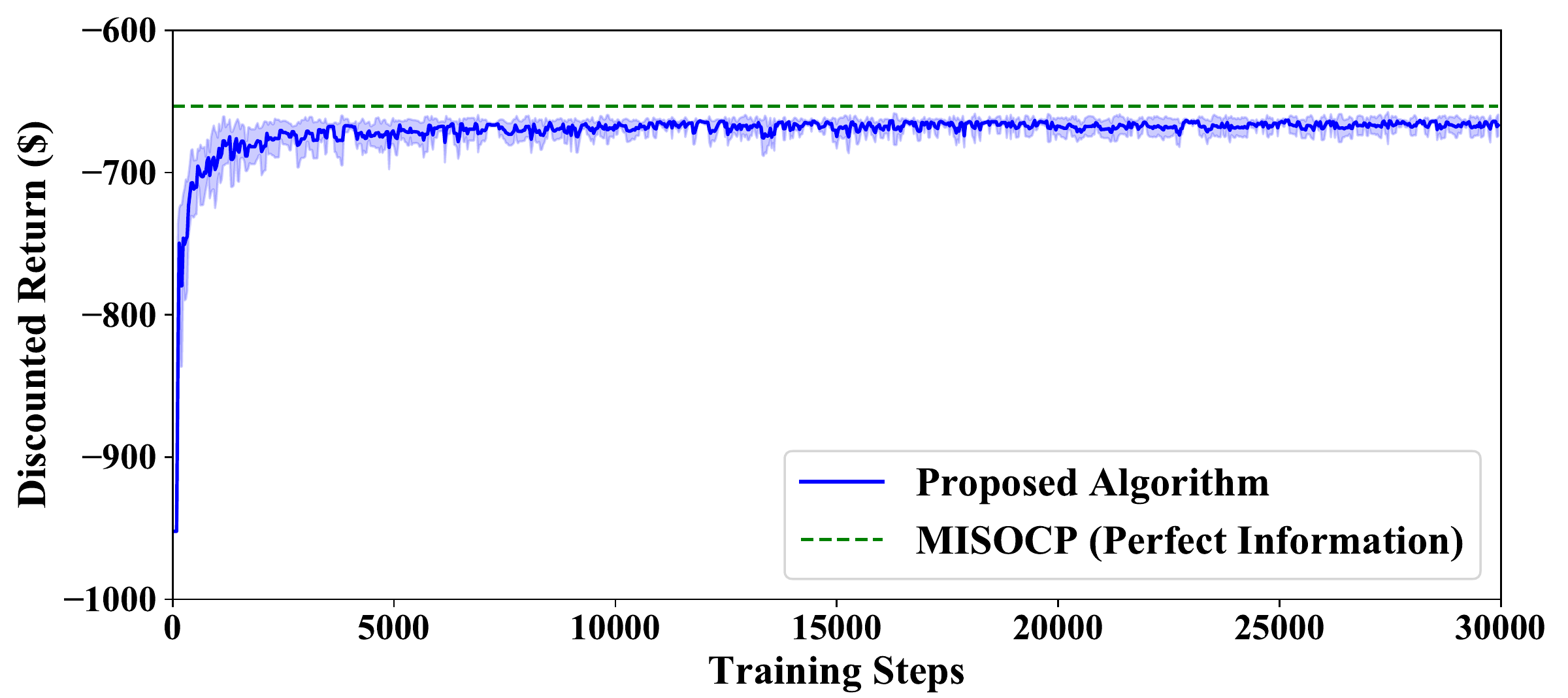}
\caption{The convergence process of the proposed algorithm. Blue solid line indicates median returns across 5 separate training runs. The dash line is the result of the dynamic programming under the perfect information. The y-axis represents the average discounted returns for the 10 validation days.
The optimal discounted returns optimized by MISOCP is -653.258 dollars.} \label{fig:MicrogridZero_Convergence}
\end{figure}

During the model training process, the network updating worker and self-play workers in Fig. \ref{fig:TrainingArch} are running parallelly on 8 CPUs simultaneously with the application of the \textit{Ray} package \cite{Ray.org}.
The proposed algorithm is trained for 30,000 steps to learn the optimal operational strategy.
In the training process, the performance of the algorithm is evaluated every 20 training steps on the selected 10-day validation data set.
Fig. \ref{fig:MicrogridZero_Convergence} illustrates the convergence process of the proposed algorithm across 5 separate runs.
On average, the training process of a single training run takes about 1 hour and 35 minutes on the computer mentioned above. 
It can be seen that the discounted returns increase rapidly with the training step increasing from 0 to 2000. 
After that the discounted returns increase slowly and finally converge around -665 dollars with small oscillations.

For each validation day, the optimal discounted returns can be obtained by using MISOCP optimization method with the assumption that the perfect RM state information is available.
That means the PV and wind power generation, and load power of each time-step is known in prior, which is not realistic in the actual online optimization process.
The obtained optimal returns are shown in Fig. \ref{fig:MicrogridZero_Convergence} as the reference to evaluate the training performance of the proposed algorithm.
As shown in Fig. \ref{fig:MicrogridZero_Convergence}, the discounted returns of the proposed algorithm approach rather than reach the optimal value.
This is because the proposed algorithm makes the online operation decisions only according to the current system state, without accurate information of the future renewable generations and electricity demand.

\subsection{The Effect of Number of Simulations per Search on the Algorithm}
\begin{figure}[t]\centering
\includegraphics[width=3.5in]{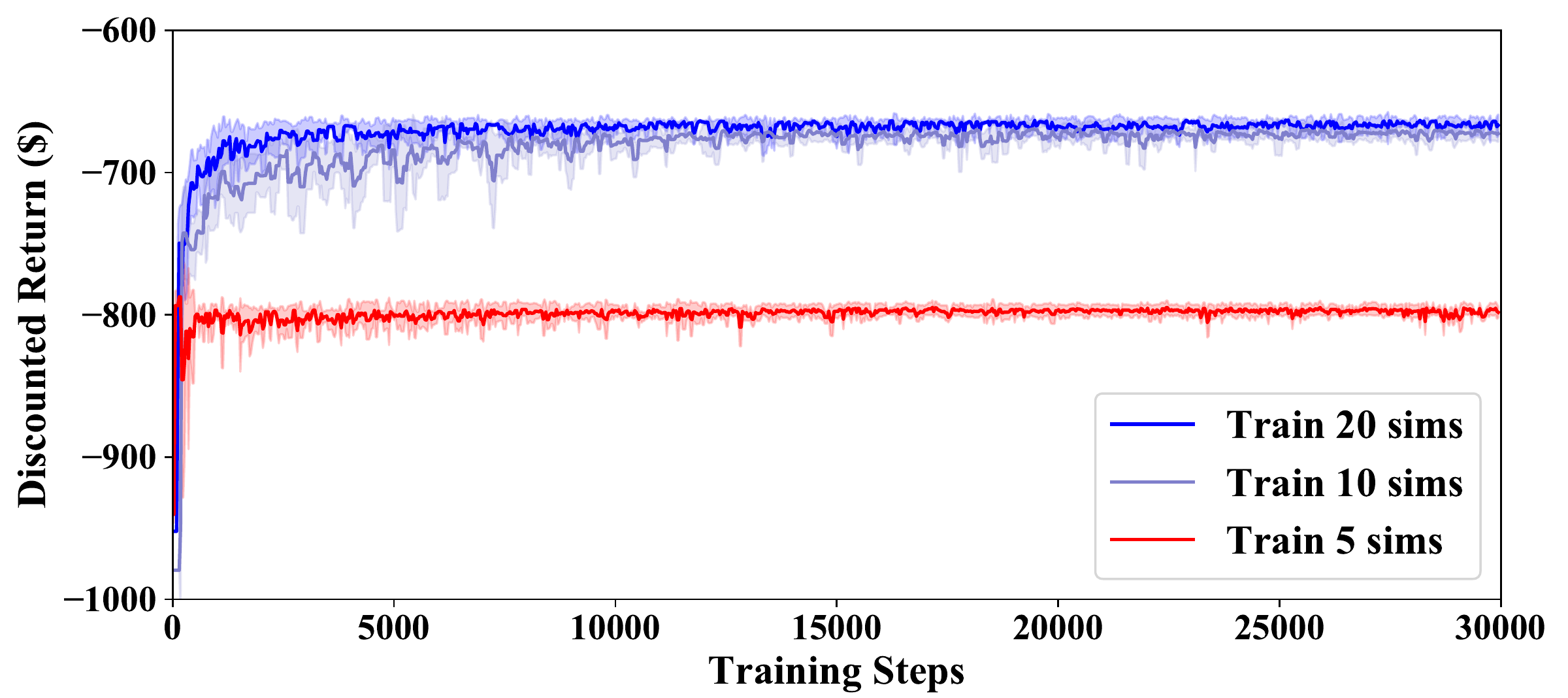}
\caption{The convergence process of the proposed algorithm at different numbers of simulations per move in tree search. Solid line indicates median returns across 5 separate training runs.} \label{fig:Diff_N_Convergence}
\end{figure}

Training with a different N (number of simulations per search) will result in a different neural network model.
In this section, the influence of $N$ on the performance of the proposed algorithm is analyzed when $N$ is set to 5, 10, and 20, respectively. 
Also, the convergence performance is evaluated across 5 separate training runs for each $N$ setting, and the results are shown in Fig. \ref{fig:Diff_N_Convergence}.
It can be found that a larger number of simulations per search will improve the performance of the algorithm.
However, the improvement becomes less obvious when $N$ increases to a certain threshold.

\subsection{Online Optimization Performance of the Proposed Algorithm}
\begin{figure}[t]\centering
\includegraphics[width=3.5in]{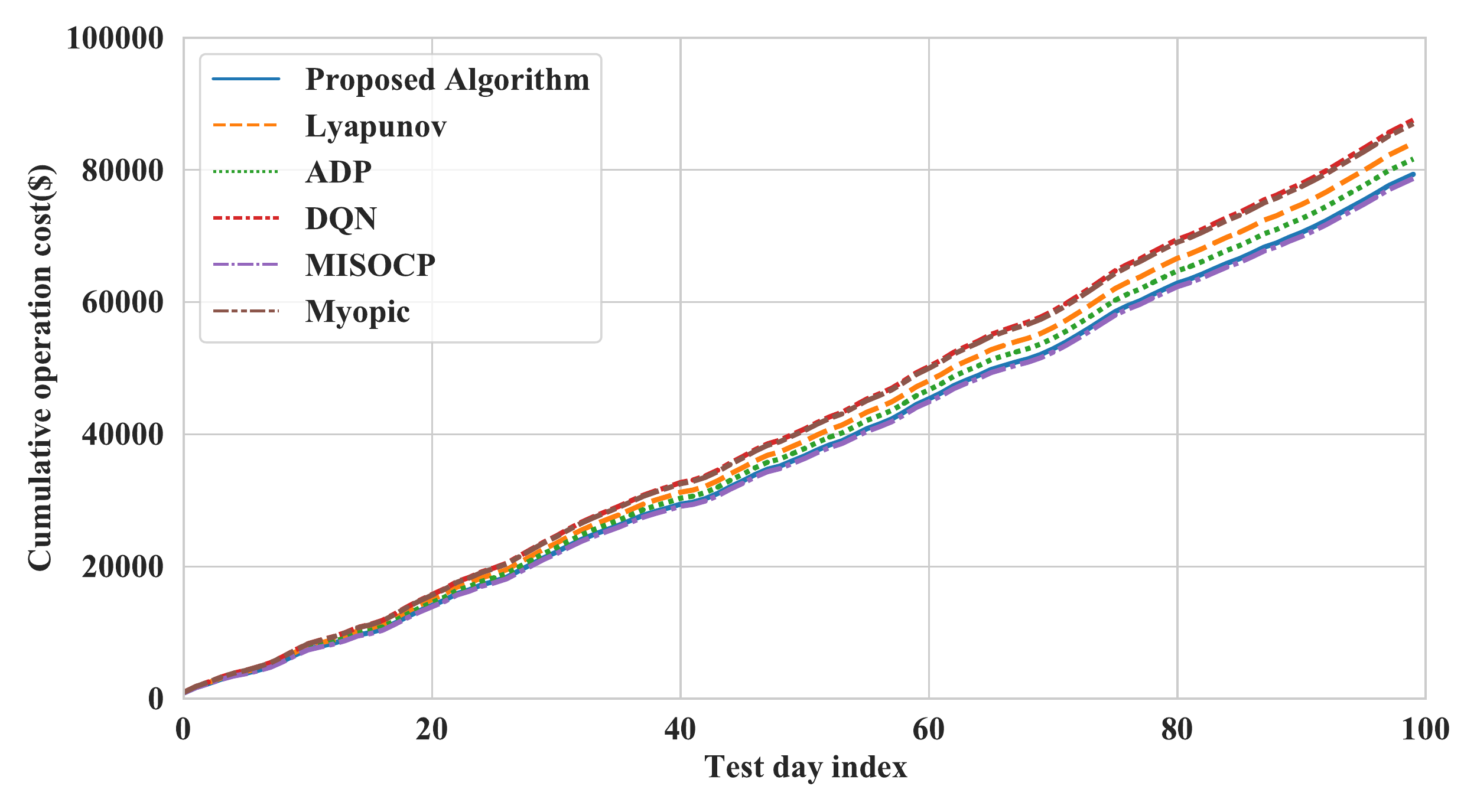}
\caption{Cumulative operation cost of the system on the testing data set.} \label{fig:Cumulative_cost}
\end{figure}

After the off-line training of the neural network model is completed, the proposed algorithm can be applied online to schedule the microgrid, as shown in \textbf{Algorithm 1}.
For the online scheduling application, the trained model in Section IV(A) is loaded and the number of simulations per search $N$ is set to 20. 
To validate the  effectiveness of the proposed algorithm, the comparison is conducted between the proposed algorithm and other state-of-art online optimization approaches, such as deep Q network (DQN), Lyapunov optimization \cite{shi2015real,wanxing2019CSEE}, ADP \cite{Hang2019TSTE}, myopic policy.
In this work, the myopic policy optimizes operation cost now, but do not explicitly consider the influence of the current decisions on the cumulative operation cost of the RM in the future.
100-day data shown in Fig. \ref{fig:TestingDataProfile} is used to test the performance, and the cumulative operational cost of the system optimized by different algorithms are calculated, as shown in Fig. \ref{fig:Cumulative_cost}.
The average daily operation cost of the RM system optimized by the proposed method, Lyapunov optimization, ADP, DQN, and myopic policy are \$ 804.08, \$ 839.65, \$ 815.79, \$ 874.56, and \$ 869.34, respectively.
Besides, the average daily operation cost optimized using the MISOCP is \$ 786.35.
The results indicate that the proposed algorithm outperforms myopic, Lyapunov optimization, ADP, and DQN algorithms in terms of the daily operation cost on average by 7.52\%, 4.24\%, 1.44\%, and 8.06\%, respectively, and yields an operation cost that closely follows the optimal cost (is only 2.25\% higher) computed by the MISOCP method under perfect information.

\begin{figure}[t]\centering
\includegraphics[width=3.5in]{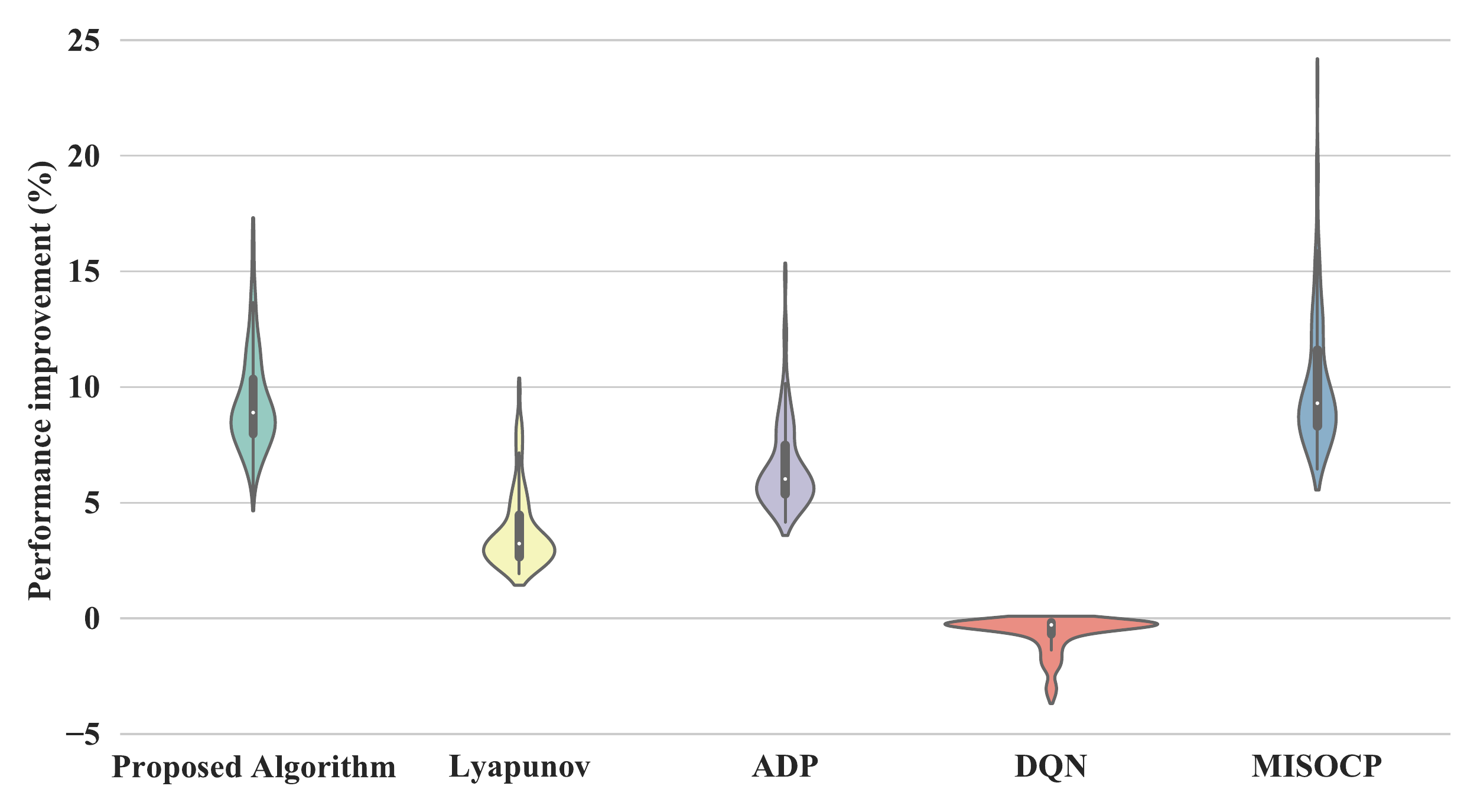}
\caption{Violin plot of the performance improvement of different online optimization methods compared to myopic method.} \label{fig:Performance_improvement}
\end{figure}

Using the result optimized by myopic policy as the baseline, the performance improvement of different methods are further evaluated, as shown in Fig. \ref{fig:Performance_improvement} and Table \ref{optimality gaps}.
Fig. \ref{fig:Performance_improvement} shows the comparison of the performance improvement of the adopted algorithms.
In Table \ref{optimality gaps}, the statistical indicators of the performance improvement of different online optimization methods are presented.
It can be found that the proposed algorithm obtains the greatest performance improvement among the online optimization algorithms.
The DQN algorithm performs worse than the Lyapunov optimization and the ADP algorithm.
We attribute the poor performance of the DQN algorithm to the low sample efficiency of the policy, and the non-stationary environment.
Although MISOCP method performs best, it is an off-line optimization algorithm and the optimal scheduling under perfect information can never be achieved since we cannot accurately forecast the future state of the microgrid system.

\begin{table}\centering
\begin{threeparttable}
\footnotesize
\newcommand{\tabincell}[2]{\begin{tabular}{@{}#1@{}}#2\end{tabular}}
\caption{The performance improvement of different optimization methods compared to myopic policy on the testing data set.}\label{optimality gaps}
\setlength{\tabcolsep}{1mm}{
\begin{tabular}
{c|c|c|c|c|c} \hline \hline
\multicolumn{2}{c|}{\tabincell{c}{$Performance$ \\ $improvement$ }} &\tabincell{c}{$Mean$ } &\tabincell{c}{$Maximum$} &\tabincell{c}{$Minimum$} &\tabincell{c}{$Standard$ \\$deviation$} \\ \hline
\multirow{4}*{Online methods} &{\tabincell{c}{Proposed \\ algorithm}} &\textbf{9.30\%}    & 16.68\%  &  5.28\% & 2.12\% \\
\cline{2-6}
&{\tabincell{c}{Lyapunov \\ optimization}}  &3.76\%    &  9.89\%  &  1.93\%  &  1.65\%  \\ 
\cline{2-6}
&{\tabincell{c}{ADP}}  &6.57\%    &  14.78\%  &  4.16\%  &  1.92\%  \\ 
\cline{2-6}
&{\tabincell{c}{DQN}}  &-0.65\%    &  -0.14\%  &  -3.45\%  &  0.77\%  \\ 
\cline{2-6} \hline
\multirow{1}*{Off-line method} &\tabincell{c}{MISOCP} &10.20\% &23.28\% &6.45\% &3.02\% \\
\hline
\hline
\end{tabular}}
\end{threeparttable}
\end{table}

\begin{figure}[t]\centering
\includegraphics[width=3.5in]{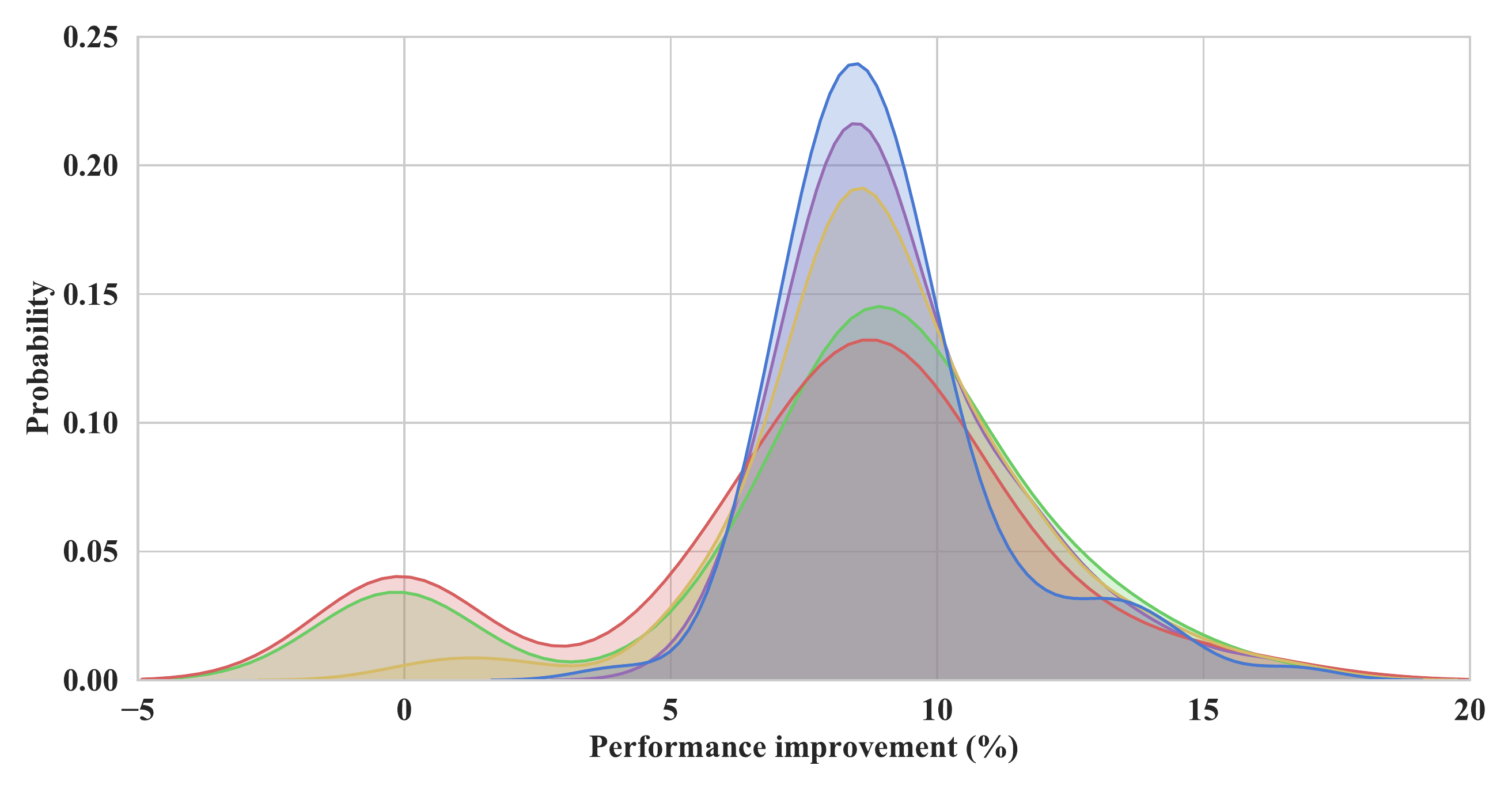}
\caption{The distribution of the performance improvement of the proposed algorithm compared to myopic policy in 5 separate runs.} \label{fig:Reproducity}
\end{figure}

\subsection{Repeatability and Feasibility of the Proposed Algorithm}
To validate the repeatability of the proposed algorithm, 5 individual off-line training runs are conducted and 5 corresponding neural network models are obtained. 
Then the online optimization performance is evaluated by applying each model on the testing data set individually.
The simulation results of the 5 individual runs are shown in Table \ref{optimality gaps2}.
It can be found that the proposed algorithm performs very stable.
\vspace{-0.5em}
\begin{table}\centering
\begin{threeparttable}
\footnotesize
\newcommand{\tabincell}[2]{\begin{tabular}{@{}#1@{}}#2\end{tabular}}
\caption{The performance improvement of the proposed algorithm compared to myopic policy in 5 individual runs.}\label{optimality gaps2}
\setlength{\tabcolsep}{3mm}{
\begin{tabular}
{c|c|c|c|c} \hline \hline
\tabincell{c}{$Performance$ \\ $improvement$ } &\tabincell{c}{$Mean$ } &\tabincell{c}{$Maximum$} &\tabincell{c}{$Minimum$} &\tabincell{c}{$Standard$ \\$deviation$} \\ \hline
\tabincell{c}{$\#1$ } &8.22\% &16.68\% &-0.17\% &3.83\% \\	\hline
\tabincell{c}{$\#2$}  &9.30\%    &  16.68\%  &  5.28\%  &  2.12\%  \\ \hline
\tabincell{c}{$\#3$}  &8.60\%    &  16.67\%  &  -0.17\%  &  3.99\%  \\ \hline
\tabincell{c}{$\#4$}  &9.04\%    &  16.68\%  &  0.33\%  &  2.59\%  \\ \hline
\tabincell{c}{$\#5$} &9.05\%    & 16.68\%  &  4.05\% & 2.04\% \\ 
\hline
\hline
\end{tabular}}
\end{threeparttable}
\end{table}

\begin{figure}[t]\centering
\includegraphics[width=3.5in]{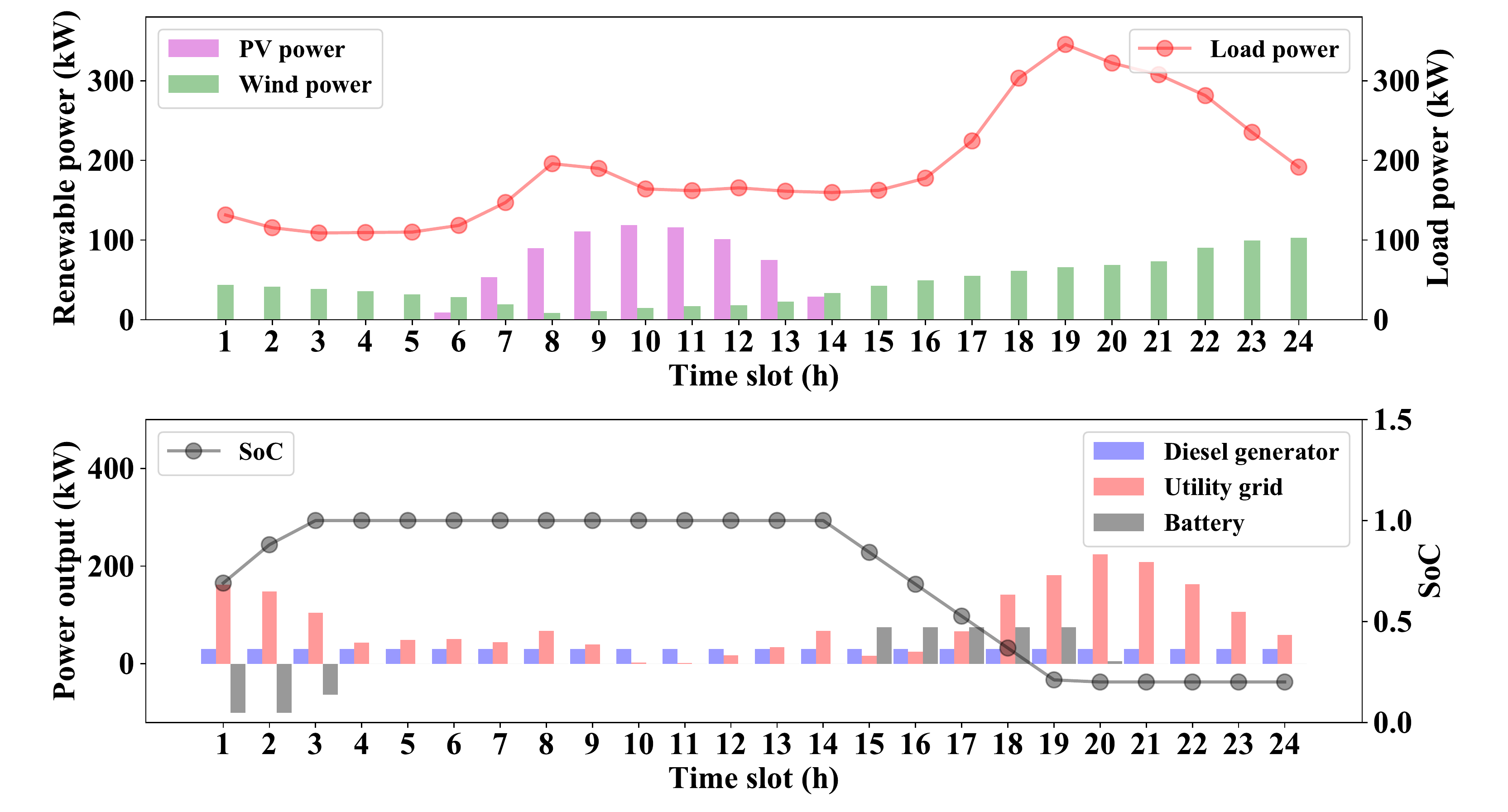}
\caption{The online energy scheduling results of the proposed algorithm.} \label{fig:SoC_PgVar}
\end{figure}

\begin{figure}[t]\centering
\includegraphics[width=3.2in]{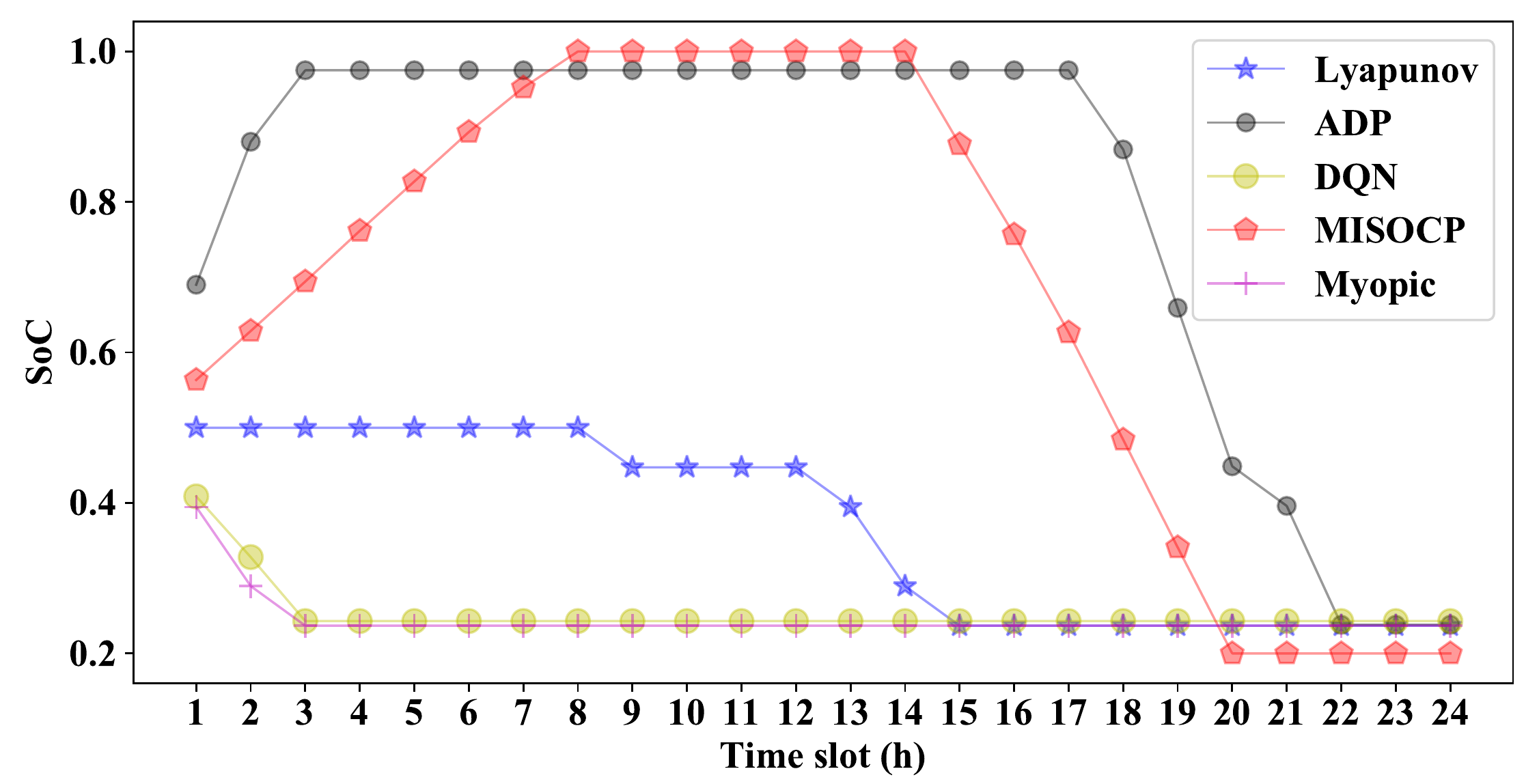}
\caption{The SoC optimized by the other algorithms.} \label{fig:SoC_AllMethods}
\end{figure}

\begin{figure}[t]\centering
\includegraphics[width=3.5in]{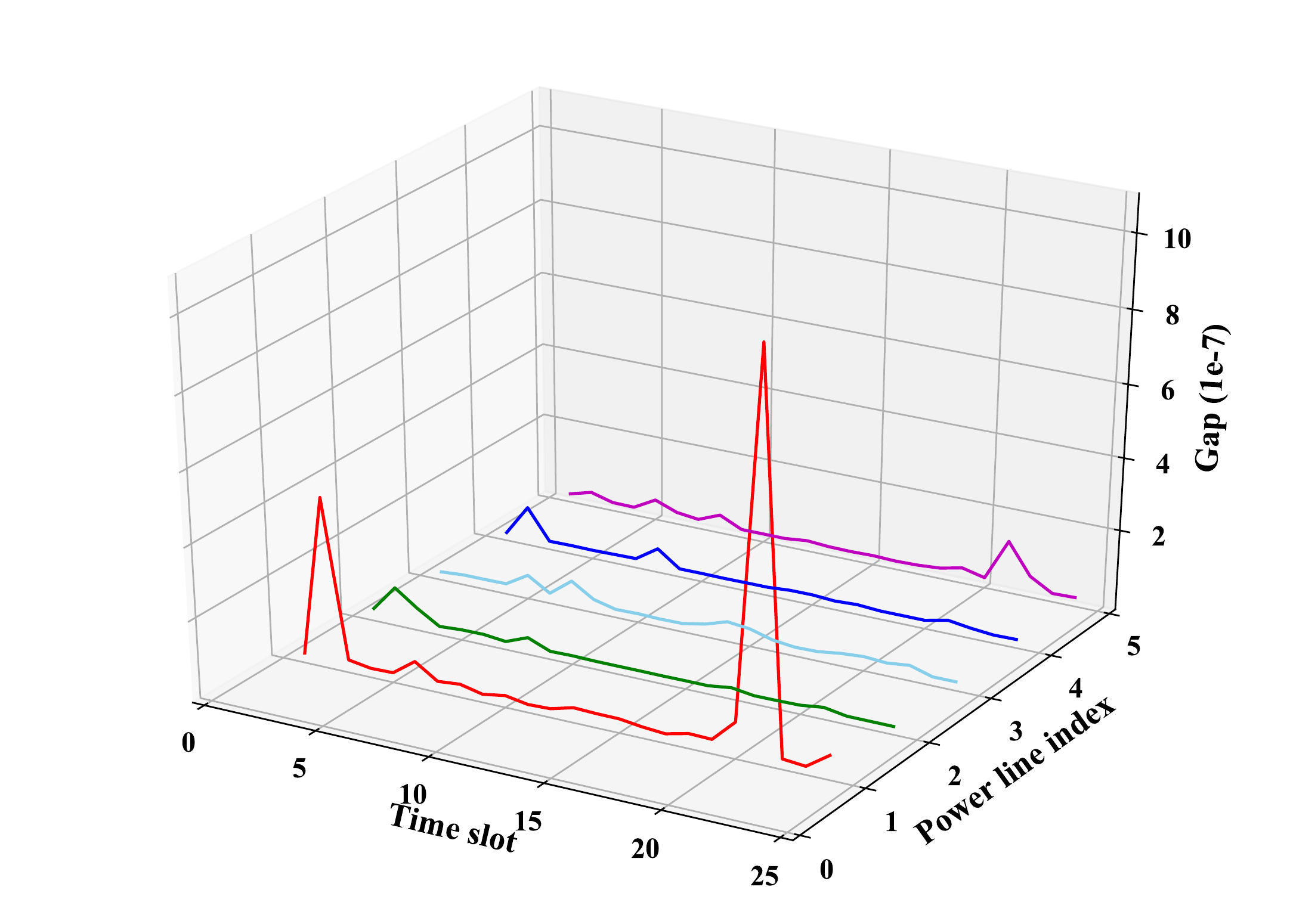}
\caption{The relaxation gaps of the power flow.} \label{fig:RelaxGap}
\end{figure}

To validate the feasibility of the decisions made by the proposed approach, the online scheduling details including the SoC pattern, power exchange between the RM and utility grid, charge/discharge power of the battery, and the power output of diesel generator are shown in Fig. \ref{fig:SoC_PgVar}. 
It can be observed that the proposed algorithm has learned to charge the battery when the electricity price is low and to discharge when the price is on-peak, and also learned to dispatch controllable generators.
Besides, the SoC patterns of the other online optimization algorithms and the optimal pattern are illustrated in Fig. \ref{fig:SoC_AllMethods}.
From the results in \ref{fig:SoC_PgVar} and Fig. \ref{fig:SoC_AllMethods}, it can be found that the proposed algorithm and the ADP method almost learned the optimal SoC pattern, while the other online optimization algorithms performs worse than the two algorithms. 
Also, the relaxation gap is calculated by $\mid (P_{ij}^2 + Q_{ij}^2)/ v_i - l_{ij} \mid$ and plotted in Fig. \ref{fig:RelaxGap}.
Note that smaller values of the gap mean better AC power flow feasibility \cite{yuan2019second}.
The maximum gap in Fig. \ref{fig:RelaxGap} is less than $10^{-6}$, which validates the effectiveness of the scheduling results in Fig. \ref{fig:SoC_PgVar}.
Finally, according to the simulation results, the proposed online optimization algorithm takes an average of 0.07s to make one single time-step scheduling, which can fulfill the time requirements of the real-time application.
The average time consumption of the Lyapunov optimization, ADP, DQN, and myopic policy to make one single time-step scheduling are 1.18s, 1.71s, 0.0038s, and 1.56s, respectively.
It can be found that the online optimization efficiency of the proposed algorithm is higher than most of the above algorithms.

\subsection{Comparison of the Proposed Algorithm and Model Predictive Control (MPC) Method}
\begin{figure}[t]\centering
\includegraphics[width=3.5in]{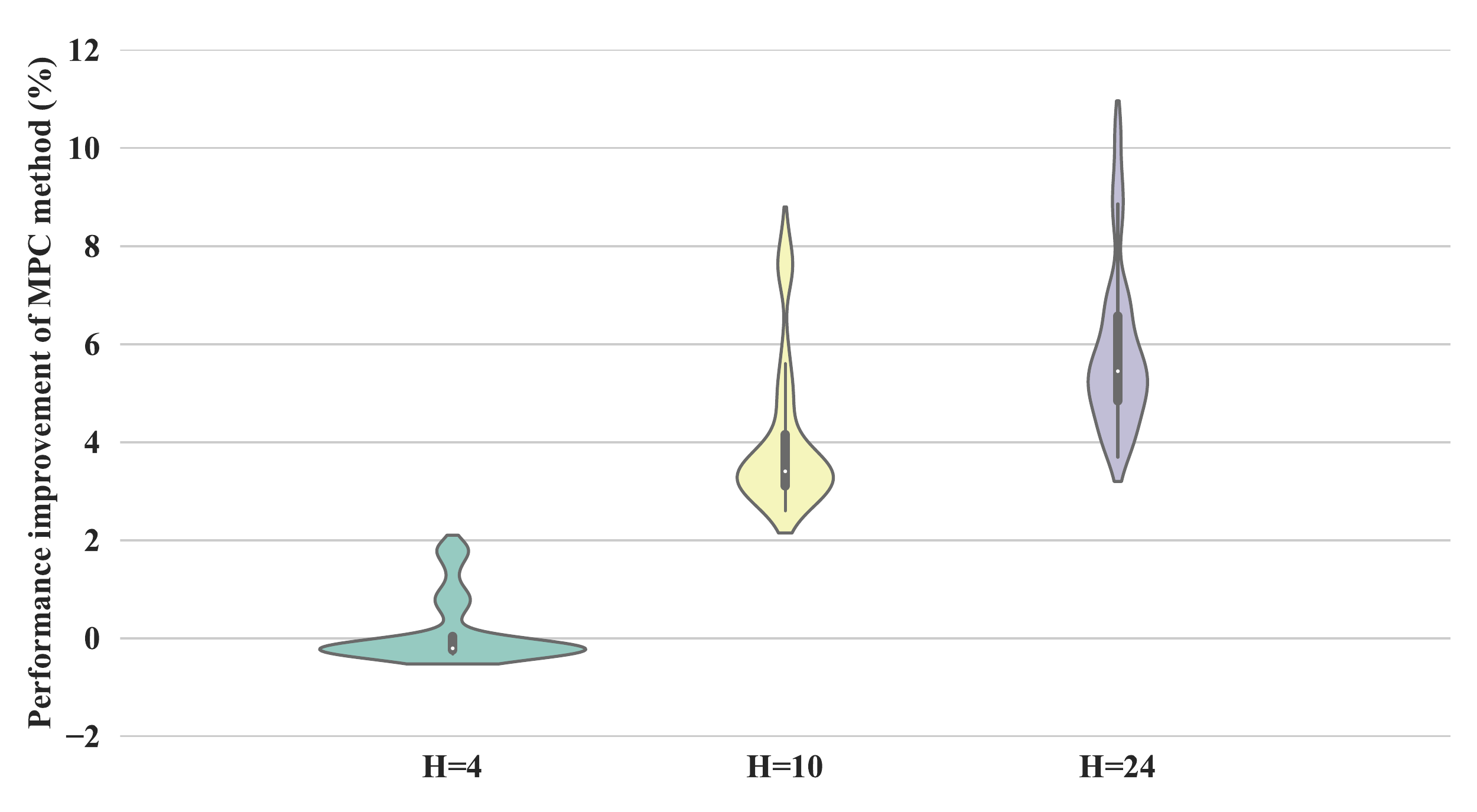}
\caption{Violin plot of the performance improvement of the MPC method compared to myopic policy.} \label{fig:MPC_online}
\end{figure}

In this section, the simulations of the online scheduling using classical MPC method are conducted and compared to the proposed algorithm.
MPC makes online decisions according to the near future prediction information from a forecast model.
For the MPC method, the prediction errors of the PV/wind power and load are set to 10\%, and 3\%, respectively.
The looking forward window ($H$) of the MPC method are set to $4h$, $10h$, and $24h$, respectively.
Fig. \ref{fig:MPC_online} shows the online optimization performance improvement of the MPC method under different $H$ values. 
As shown in Fig. \ref{fig:MPC_online}, the average online optimization performance improvements of the MPC method compared to myopic policy are 0.13\%, 4.03\%, and 5.84\%, respectively.
It can be found that, the performance improvement of the MPC method increases with the increase of $H$.
However, it still underperforms the proposed algorithm (9.30\% as shown in Table \ref{optimality gaps}), even if the forecasting information of PV/wind/load power in the next 24 hours are utilized.
Besides, the average time consumption of the MPC method ($H = 24 h$) to make one single time-step scheduling is 3.83s, which is much longer than the time needed by the proposed algorithm.

\section{Conclusion} \label{Conclusion}
This paper investigates the optimization algorithm for the online optimization of a RM under uncertainties, which is based on a newly developed MB-DRL algorithm, MuZero \cite{schrittwieser2019mastering}. 
It combines a Monte-Carlo tree search method with a learned neural network model which consists of representation network, dynamic network, and prediction network. 
To deal with the huge decision space and constraints in the RM model, a framework was proposed to combine the MB-DRL method with the SOCP technique.
With the consideration of the characteristics of the optimization problem in this work, the representation network is redesigned to adopt LSTM units to extract features from historical RM state data and feed into a fully connected neural network.
The parameters of these networks are updated by gradient descent using the operation data generated by the self-play mechanism during the training process.
Numerical simulation results show that the proposed approach can learn to operate the microgrid by itself from data of self-play, and can make online decisions without dependence on renewable and load power prediction information from forecast models.
Besides, the proposed algorithm outperforms the state-of-the-art online optimization methods, including the DQN, Lyapunov optimization, ADP, myopic policy, and MPC method.

\bibliographystyle{IEEEtran}
\bibliography{Zero_microgrid_OnlineOptimization}
\vspace{1em}
\section*{Appendix} \label{AppendixA}
\vspace{0.5em}
\begin{breakablealgorithm}   \caption{The training process of the RM online optimization algorithm.}\label{alg:MicrogridZero}
\small
\begin{algorithmic}[1]
    \State Initialize the neural networks of the model; set the total number of training steps $N_T$, number of simulations per search $N$, discount factor $\gamma$, batch size $B$, replay buffer size $W$, unroll steps $K$, bootstrapping steps $n$, ratio of self-play speed to training speed $\upsilon_s$, and other hyperparameters. Initialize the training step index $n_s = 0$.
    \State Initialize self-play workers, training worker, replay buffer worker, and shared storage workers, then launch workers. \Comment {\textit{{\footnotesize Self-play workers run on several CPUs in parallel.}}} 
	\While {$n_s \leq N_T$}:
		\begin{enumerate}
			\item \emph{D} $\leftarrow$ \textit{\MakeUppercase{Self-Play()}}. \Comment {\textit{{\footnotesize Generate RM operation data by self-play workers.}}}
			\item \textit{\MakeUppercase{Replay Buffer.ADD}}(\emph{D}). \Comment {\textit{{\footnotesize Save the generated operation data in the replay buffer.}}}
			\item ($\theta, \vartheta, \phi$) $\leftarrow$ \textit{SHARED STORAGE.LATEST}(). \Comment {\textit{{\footnotesize Get latest model parameters.}}}
			\item \emph{B} $\leftarrow$ \textit{REPLAY BUFFER.SAMPLE}(). \Comment {\textit{\footnotesize {Sample batch data \emph{B} from the replay buffer worker}}}.
			\item ($\theta_{new}, \vartheta_{new}, \phi_{new}$) $\leftarrow$ \textit{NETWORK TRAINING}($\theta, \vartheta, \phi$, $B$, $n_s$) \Comment {\textit{\footnotesize {Update the latest model}}}.
			\item \textit{SHARED STORAGE.ADD}($\theta_{new}, \vartheta_{new}, \phi_{new}$) \Comment {\textit{\footnotesize {Save the updated model parameters in the shared storage worker}}}.
			\item $n_s = n_s + 1$
		\end{enumerate}
	\EndWhile 
	\State Return the latest neural network model.
\end{algorithmic}
\end{breakablealgorithm}

\begin{breakablealgorithm}   \caption{SELF-PLAY}\label{alg:Self-Play}
\small
\begin{algorithmic}[1]
\Require  
	  The model;      
      the PV power, wind power, and load power scenario;
      the training index $n_s$; 
    \Ensure  
      The generated RM operation data. \Comment{\textit{\footnotesize {The procedure includes 3 parts: get the latest model from the shared storage, play game, save game to replay buffer.}}}
	\State Load ($\theta, \vartheta, \phi$). \Comment{\textit{\footnotesize {Get the latest model from the shared storage}}}
\State Randomly select a day from the training data set and load the data. \Comment{\textit{\footnotesize {PV/wind sequence, load sequence, and price sequence.}}} 
    \State Reset the RM simulation environment, and get the initial state $s_0$. 
    \For {$t = {0, 1, 2, \cdots, T-1}$}: \Comment{\textit{\footnotesize {$T = 24$ in this work.}}}
			\State Internal state $\hat s_t \leftarrow h_{\theta}(s_t,s_{t-1}, \cdots, s_0)$  \Comment{\textit{\footnotesize {(\ref{EQ29})}}}.
			\State Create root node $\gamma_{root}$ with state $\hat s_t$.
			\State Run \textbf{MCTS($\gamma_{root}$, $\theta, \vartheta, \phi$, $\chi$)}. \Comment{\textit{\footnotesize {$\chi$ is the decision history}}}.
			\State Select $P_b(t)$ that leads to the most visited child of root node.
			\State Overcharge/overdischarge check. \Comment{\textit{\footnotesize {(33)}}}.
			\State Solve the OPF sub-problem to get the remaining decisions.
			\State Apply the optimal decision to the RM, and get actual reward 
			\Statex \quad \ $r_t$ and the next state $s_{t+1}$. \Comment{\textit{\footnotesize {(\ref{EQ18}), (\ref{EQ28})}}}.
			\State Store search statistics. \Comment{\textit{\footnotesize {visit counts, root value, etc.}}}
	\EndFor
	\State If the ratio of replay buffer data size to the shared storage data size is greater than $\upsilon_s$, pause the self-play worker for 0.5 seconds.
	\State Return generated RM operation data. \Comment{\textit{\footnotesize {($s_0, x_0, r_0, v_0, s_{1}$), ($s_1, x_1, r_1, v_1, s_{2}$), $\cdots$}}}
	\State
\Function{\textbf{MCTS}}{$\gamma_{root}$, $\theta, \vartheta, \phi$, $\chi$}
	\For {$n_{sim} \in {1, 2, \cdots, N}$}
	\State $node$ $\leftarrow$ $\gamma_{root}$.
	\State history = $\chi$, $SearchPath$ = [$node$].
	\While {$node$ is not leaf node}
	\State {Select the child node with highest UCB score, and get 
	\Statex \qquad \qquad \ the charge/discharge decision $\hat x_k$.  \Comment{\textit{\footnotesize {(\ref{EQ34}).}}}}
	\State history.append($\hat x^k$)
	\State $node$ $\leftarrow$ new child node.
	\State $SearchPath$.append($node$).
	\EndWhile
	\State $parent$ $\leftarrow$ $SearchPath$[-2].
	\State Using the dynamic network, compute the reward $\hat r^k$
	\Statex \qquad \quad and the next internal state $\hat s^{k+1}$ after taking decision 
	\Statex \qquad \quad history[-1] from $parent$. \Comment{\textit{\footnotesize {(\ref{EQ35})}}}
	\State Using the prediction network, compute the policy $p^{k+1}$
	\Statex \qquad \quad and value $v^{k+1}$. \Comment{\textit{\footnotesize {(\ref{EQ36})}}}
	\State According to $\hat s^{k+1}$, $\hat r^k$, $p^{k+1}$, and $v^{k+1}$, update parameters 
	\Statex \qquad \quad of the $node$ and expand $node$.
	\For {$node$ in reversed($SearchPath$)}
	\State value = $v^{k+1}$
	\State $node.value$ = $node.value$ + value.
	\State $node.visit\_count$ = $node.visit\_count$ +1.
	\State value = $node.reward$ + $\gamma$ * value. \Comment{\textit{\footnotesize {(\ref{EQ38}) -(\ref{EQ40})}}}
	\EndFor				
	\EndFor
	\EndFunction

\end{algorithmic}
\end{breakablealgorithm}

\begin{breakablealgorithm}   \caption{NETWORK TRAINING.}\label{alg:TRAINING}
\small
\begin{algorithmic}[1]
	\Require  
      The model parameter {$\theta, \vartheta, \phi$, $B$};  
      the training index $n_s$. 
    \Ensure  
      The updated model.     
    
    \State Set the learning rate and the optimizer. \Comment{\textit{\footnotesize {Constant learning rate in this paper.}}} 
	\State Load ($\theta, \vartheta, \phi$). \Comment{\textit{\footnotesize {Get the latest model from the shared storage}}}
	\State Get the training data \emph{B}.
	\State {According to the batch data \emph{B}, compute the evaluation of the value, reward, policy, and internal state using the model ($\theta, \vartheta, \phi$).}  
	\State Get the target from the batch \emph{B}, then calculate the cross entropy loss using {(\ref{EQ41})} and {(\ref{EQ42})}.
	\State Update the model parameters using the Adam optimizer.
	\State If the ratio of replay buffer data size to the shared storage data size is less than $\upsilon_s$, pause the network training worker for 0.5 seconds.
	\State Return the updated model.  
\end{algorithmic}
\end{breakablealgorithm}

\end{document}